\documentclass[10pt, journal, a4paper, final, nofonttune]{IEEEtran}
\usepackage{amssymb,amsmath,amsthm,epsfig,graphics,float,subfigure, mhchem}

\begin{document}
\title{Electrochemical characterisation of ionic dynamics resulting from spin conversion of water isomers}
\author{Serge Kernbach\\[3mm]
\small CYBRES GmbH, Research Center of Advanced Robotics and Environmental Science, \\
\small Melunerstr. 40, 70569 Stuttgart, Germany, \emph{serge.kernbach@cybertronica.de.com}\\
\vspace{-5mm}}

\date{}
\maketitle
\thispagestyle{empty}

\begin{abstract}
Para- and ortho- isomers of water have different chemical and physical properties. Excitations by magnetic field, laser emission or hydrodynamic cavitation are reported to change energetic levels and spin configurations of water molecules that in turn change macroscopically measurable properties of aqueous solutions. Similar scheme is also explored for dissolved molecular oxygen, where physical excitations form singlet oxygen with different spin configurations and generate a long chain of ionic and free-radical reactions. This work utilizes electrochemical impedance spectroscopy (EIS) to characterize ionic dynamics of proposed spin conversion methods applied to dissolving of carbon dioxide \ce{CO_2} and hydrogen peroxide \ce{H_2O_2} in pure water excited by fluctuating weak magnetic field in $\mu T$ range. Measurement results demonstrate different ionic reactivities and surface tension effects triggered by excitations at $10^{-8}$J/mL. The \ce{CO_2}- and \ce{O_2}-related reaction pathways are well distinguishable by EIS. Control experiments without \ce{CO_2}/\ce{H_2O_2} input show no significant effects. Dynamics of electrochemical impedances and temperature of fluids indicates anomalous quasi-periodical fluctuations pointing to possible carbonate-induced cyclic reactions or cyclical spin conversion processes. This approach can underlay development of affordable sensors operating with spin conversion technologies.
\end{abstract}

\section{Introduction}

Para- and ortho- isomers \cite{Tikhonov02} of water have different physical and chemical properties: chemical reactivity \cite{Kilaj18}, behaviour in electric field \cite{Horke:192659}, magnetic moment \cite{doi:10.1126/science.1200433}, viscosity and surface tension \cite{Pershin15Biophysics} and several other properties. Research on spin isomers enables exploring ice-like structures in liquid water \cite{Pershin09}, previously unclear effects of treatments in magnetic fields \cite{Vaskina2020} and hydrodynamic cavitation \cite{Pershin12yeast}. Recent publications indicated an important role of spin isomers in biological organisms: purple bacteria \cite{Pishchalnikov13}, yeast cells \cite{Pershin12yeast}, photosynthesis complex \cite{Pershin12Phytosyntesis}, erythrocytes \cite{Pershin09} as well as more general spin-related mechanics in quantum biology \cite{Cao20}, \cite{quantum3010006}.

Experiments reported deviations from equilibrium ortho/para ratio 3:1 towards non-equilibrium ratios, for instance in ice \cite{Buntkowsky21}. Pershin and co-authors demonstrated achieving the non-equilibrium ratio of 1:1 in bulk water at room temperature \cite{Pershin15Biophysics}, \cite{Pershin12Phytosyntesis}, and opened discussions about weak energetic excitations with minimal energetic level at $\sim 10^{-8}$ eV (about $\sim 10^{-27}$J) \cite{Pershin08}, \cite{Pershin15Nano}, typically mechanical, laser and EM treatments. There are several attempts of theoretical explanation \cite{Konyukhov11} for such spin conversion mechanisms. Spin isomers are expected to generate periodical and quasiperiodical oscillations with different periods in biochemical reactions \cite{Morre15} and biological organisms \cite{Drozdov14}. The topic of spin conversion/reorientation under energy flux is not new, for instance, similar mechanisms in singlet oxygen are well investigated in application to chemical and biological processes \cite{Wayne69},\cite{Shcherbakov20}.

Detection of spin isomers is performed by different physical methods such as Raman spectroscopy \cite{Pershin12Phytosyntesis}, far-infrared spectroscopy and nuclear magnetic resonance \cite{Buntkowsky21}, THz absorption spectroscopy \cite{doi:10.1063/1.2357412} and others. However, since the spin conversion influences various chemical and physical parameters, indirect detection and real-time monitoring can be performed in different ways. For instance, \cite{Pyzhyanova18} uses refractive index for measuring the level of singlet \ce{^1O_2}, \cite{Spangenberg21} applied UV spectroscopy for indirect detection of contaminants in water. We expect that also electrochemical methods, in particular, electrochemical impedance spectroscopy (EIS) and fast EIS \cite{boskoski2017fast} can serve for detection and characterization purposes. Additionally, measurements of physical parameters such as kinematic viscosity (surface tension), refractive index or UV absorption spectra can provide complimentary data.

This work explores the spin conversion hypothesis from \cite{Pershin15Biophysics}, \cite{Pershin12Phytosyntesis} applied to \ce{CO_2} dissolving process, i.e. hypothetical reactions \ce{CO_2 + p-H_2O} and \ce{CO_2 + o-H_2O} in pure water are expected to demonstrate different electrochemical reactivity. Kinetics of \ce{CO_2} dissolving is well investigated \cite{Mitchell09}, \cite{Capobianco14} that leads to accumulation of \ce{H_3O^+}, \ce{HCO_3^-} and \ce{CO_3^{2-}} ions in six different time scales, and continuous increasing of conductivity up to the saturation level. Additionally, we also test the pathway of reactive oxygen species (ROS) under external excitation that forms, among others, singlet oxygen \ce{^1O_2} and hydrogen peroxide \ce{H_2O_2} with different free radicals and ions \cite{Belovolosova20}. Thus, imposing energy excitation, primarily the weak magnetic field in $\mu T$ range, and excluding or minimizing other factors, we expect a specific dynamics of ionic production.

While optical measurements are a well-established approach, EIS \cite{Wang21} provides several advantages: less expensive equipment, operations in frequency or in time domains, measuring physical effects such as ionic mobility, sensitivity to weak ionic changes, large number of electrochemical models that enable analytical methods. Therefore, this work is focused on exploration of EIS as an additional tool for rapid or in-field characterization of ionic processes resulting from spin conversion, which has been already approached in other works.

\section{Secondary factors, methods and setup}
\label{sec:secFactors}

\ce{CO_2} dissolving in water forms carbonic acid \ce{H_2CO_3}
\begin{equation}
\label{eq:carbonicAcid}
H_2O + CO_{2(aq)} \rightleftharpoons H_2CO_3
\end{equation}
that dissociates to \ce{H_3O^+} and \ce{HCO_3^-}.  The ion \ce{OH^-} is required for dissociation of \ce{H_2CO_3} that is delivered by the auto-ionization process \cite{Capobianco14}. The equilibrium reaction (\ref{eq:carbonicAcid}) is obeying Le Chatelier's principle; the forward reaction is exothermic, an increase in temperature causes reverse reaction. Following \cite{Capobianco14}, the ionic dynamics of
\begin{equation}
\label{eq:carbonicAcidIons}
mH_2O + CO_2 \rightleftharpoons H_3O^+ + HCO_3^- + (m-2)H_2O
\end{equation}
is monotonic with short-term oscillation in initial phase. Bicarbonate $HCO_3^-$ further dissociates and forms \ce{CO_3^{2-}}
\begin{equation}
\label{eq:carbonicAcidIonsFurther}
HCO_3^- \rightleftharpoons  H_3O^+ + CO_3^{2-}.
\end{equation}
According to \cite{Mitchell09}, \ce{H_2CO_3} and \ce{HCO_3^-} have an approximately linear evolution in time, \ce{CO_3^{2-}} is varying quadratically (this work expressed six time scales for each of ionic species).

Electric conductivity and temperature have the following dependency
\begin{equation}
\label{eq:t}
EC_t=EC_{25}[1+a(t-25)],
\end{equation}
where $a$ varies between 0.0191 and 0.025, $EC_t$ is electrical conductivity at temperature $t$, $EC_{25}$ is electrical conductivity at 25C \cite{Hayashi2004}. For instance, an increasing of water temperature of $\Delta 0.1$C at 0.5MOhm causes the drop of impedance about $\Delta 1$kOhm close to 25C. To explore temperature effects, active thermostats are not used used in experiments (only passive thermal stabilization), the temperature of fluids and environments is recorded and analysed based on (\ref{eq:t}).

There are two arguments in discussion \cite{Pershin15Biophysics} about changes of conductivity of para-/ortho- isomers in bulk water excited by weak magnetic field. The first argument is related to a higher reactivity of para isomers, as demonstrated in \cite{Kilaj18} for
\begin{equation}
o-/p-H_2O + N_2H^+ \rightarrow N_2 + H_3O^+,
\end{equation}
where \ce{p-H_2O} has about 23(9)\% higher reactivity and produces more ionic products. The second argument is related to viscosity (as well as surface tension) and ionic mobility, as demonstrated in \cite{Pershin09}, \cite{Pershin15Nano}. Electrochemical impedance depends on the number of ions and their mobility, both mechanisms are expected to affect the measurements. Since changes of reactivity in (\ref{eq:carbonicAcid}) are electrochemically non-detectable, EIS can measure the ionic products of (\ref{eq:carbonicAcidIons}). Separate setup with open-surface electrodes explores additional effects related to surface tension in electrochemical measurements, see Fig. \ref{fig:viscosity}.

The proposed deviation from 3:1 ortho-/para- ratio towards non-equilibrium 1:1 ratio increases reactivity of (\ref{eq:carbonicAcid}), which results in different  slopes of impedance dynamics, see Figs. \ref{fig:global5} and \ref{fig:global6}. However, the equilibrium equations (\ref{eq:carbonicAcid})-(\ref{eq:carbonicAcidIonsFurther}) can increase or decrease impedance as forwards/reverse reactions. Increasing of temperature will decrease the impedance ($-\Delta Im: + \Delta t$) by (\ref{eq:t}), and, at the same time, will increase the impedance ($+\Delta Im: Rev$) by reverse (\ref{eq:carbonicAcid})-(\ref{eq:carbonicAcidIonsFurther}). Due to a switching character of forwards/reverse (\ref{eq:carbonicAcid})-(\ref{eq:carbonicAcidIonsFurther}) and temperature variations, we can expect sharp changes between up and down trends of impedance dynamics. Such points can be considered as electrochemical markers, see Sec. \ref{sec:elecrochemicalMarkers}.

Beside the discussed ortho-/para- conversion, several other effects contribute to ionic properties of pure water: 1) pathway of dissolved molecular oxygen, building of hydrogen peroxide \ce{H_2O_2} and singlet oxygen \ce{^1O_2} \cite{Shcherbakov20}; 2) near surface effects: absorption of hydronium ions \ce{H_3O^+} \cite{Lee20}, charge separation in EZ layer at air-water interface \cite{Chai09} and structuring of EZ water \cite{PMID:34855917}; 3) acoustic and mechanical influences \cite{Ayrapetyan99},  \cite{ijms21218033}; 4) dissociation and self-ionization \cite{2001Sci291.2121G}.

The oxygen pathway is the most important factor since it represents a molecular mechanism of converting physical influences into the chemical form. First of all, the formation of reactive oxygen species (ROS) is a natural process that occurs continuously due to the Earth's background radiation -- every second in 1 $cm^3$ of water there are $10^4$ formations of ionized states and $3\cdot{10^4}$ -- of excited states, and \ce{H_2O_2} is present in all natural waters \cite{Belovolosova20}. We denote this as natural oxygen pathway, the main question is about its influence on measurements since such excited states are extremely short. Secondly, optical excitations at 0.15 J/mL (1264nm) lead to change of energy level from triplet to singlet state and to different spin configurations of molecular oxygen \cite{Gudkov11}. Experiments with 650nm and 1270nm laser emission and 1-7 T constant magnetic field demonstrated an increase of singlet \ce{^1O_2} in distilled water \cite{Pyzhyanova18}, \cite{Shcherbakov20}. Adding a small amount of \ce{H_2O_2} can essentially stimulate ROS reactions, and generate even cyclical ones \cite{Voeikov10} -- we denote this as induced oxygen pathway.

We can sketch this pathway by following \cite{Voeikov06}, \cite{Voeikov07}, \cite{Caer11}, \cite{Belovolosova20}. Liquid water has a high absorption in a wide wavelength range (beside visible part) and thus has different schemes of transforming energy into thermal and chemical forms. In the first step, the energy flux forms ionized and excited water molecules \ce{H_2O^+}, \ce{H_2O^*}, and sub-excitations electrons \ce{e^-}. In the second step, electrons are solvated, \ce{H_2O^+} and \ce{H_2O^*} are dissolved
\begin{equation}
\label{eq:secondStep}
H_2O^* \rightarrow H^\bullet + HO^\bullet, ~~~H_2O^+ \rightarrow H_3O^+ + HO^\bullet.
\end{equation}
Based on work of Domrachev, Voeikov in \cite{Voeikov06} discussed an interesting scheme of homolytic dissociation of the \ce{HO-H} bond under low density energy flux
\begin{eqnarray}
(H_2O)_n (H_2O...H-|-OH)(H_2O)_m + E \rightarrow \nonumber \\ \rightarrow (H_2O)_{n+1} (H^\bullet)+(OH^\bullet) (H_2O)_m
\end{eqnarray}
as a mechanically excited polymeric entity that produces \ce{H^$\bullet$} and \ce{OH^{$\bullet$}} radicals and then recombines them back to water with 5.2eV energy
\begin{equation}
\label{eq:domrachev}
H^\bullet + OH^\bullet \rightleftharpoons H_2O + e^-_{aq}.
\end{equation}
Pollack \cite{Chai09} introduces another mechanism of energy conversion based on properties of EZ water
\begin{equation}
\label{eq:polack}
(H_3O_2)^-_n \rightleftharpoons n(H_2O+HO^\bullet + e^-_{aq}).
\end{equation}
The third step finally leads to various chains of chemical reactions depending on energy flow, existing molecules in fluids, $pH$ and other factors, e.g.
\begin{eqnarray}
\label{eq:thirdStep}
e^-_{aq} + O_2 \rightarrow O_2^{\bullet-},~~H^\bullet + O_2 \rightarrow HO_2^{\bullet},~~HO_2^{\bullet} \rightleftharpoons H^+ + O_2^{\bullet-}. \nonumber
\end{eqnarray}
Further dynamics of \ce{HO^$\bullet$}, \ce{HO_2^{$\bullet$}} and \ce{O_2^{$\bullet$-}} depends on $pH$, and produces \ce{H_2O_2} and singlet oxygen
\begin{equation}
HO^\bullet+ HO^\bullet \rightarrow H_2O_2, ~~~HO_2^{\bullet} + HO_2^{\bullet} \rightarrow H_2O_2+ ({}^1 O_2^{*}, hv).\nonumber
\end{equation}
\ce{H_2O_2} will either slowly dissociate in the ionic way
\begin{eqnarray}
\label{eq:ionic}
H_2O_2 \rightleftharpoons HO_2^- + H^+, ~~HO_2^- \rightleftharpoons O_2^{2-} + H^+
\end{eqnarray}
or with a growing concentration and in presence of \ce{O_2^{$\bullet$-}} in the fast radical way
\begin{eqnarray}
H_2O_2 + O_2^{\bullet-} \rightarrow OH^- + OH^{\bullet} + O_2, \nonumber \\
OH^{\bullet} + H_2O_2 \rightarrow HO_2^{\bullet} + H_2O.
\end{eqnarray}
\ce{OH^$\bullet$} reacts with ions from \ce{CO_2} pathway \cite{Belovolosova20}:
\begin{eqnarray}
OH^{\bullet} + HCO_3^- \rightarrow CO_3^{\bullet-} + H_2O, \nonumber\\
CO_3^{\bullet-} + OH^{\bullet} \rightarrow CO_2 + HO_2^-, \nonumber \\
CO_3^{\bullet-} + H_2O_2 \rightarrow HCO_3^- + HO_2^{\bullet},
\end{eqnarray}
here we expect an oscillating ionic dynamics around \ce{HCO_3^-} as reported e.g. in \cite{Voeikov10}. Adding \ce{H_2O_2} to liquid water forms different ions
\begin{eqnarray}
H_2O_2 + H_2O \rightleftharpoons H_3O^+ + HO_2^-,
\end{eqnarray}
that without energy flux uses the scheme (\ref{eq:ionic}) for a slow ionic dissociation and demonstrates a degradation of impedance. Chemical reactions for generating \ce{H_2O_2} (\ref{eq:secondStep})-(\ref{eq:thirdStep}) are based on radicals, here no essential or fast changes of impedance are expected. The singlet oxygen \ce{^1O_2} can represent another spin-reorientation mechanism that will be also shortly explored.

For testing the main hypothesis, it needs to exclude or minimize these secondary factors. Thus, the \ce{CO_2}-controlled atmosphere between 700-12000ppm is used with the gas sensor (Senserion SCD4x), ionic saturation in \ce{CO_2}-enriched atmosphere at 12000ppm (saturated value) occurs very fast within about 10 minutes, see Fig. \ref{fig:fastCO20}. Pure water (bidistilled with 0.05$\mu S/cm$ and distilled with 1$\mu S/cm$, 10-15 ml) is used, measurements are performed in closed thermo-insulating boxes (neopor, 5 cm wall thickness with 5 litres of water inside to create a thermal inertia) in darkness. In experimental methodology, several setups are explored: without access to atmosphere, \ce{CO_2}-enriched or normal atmosphere. Water samples in initial attempts were degassed by repeated sonication under light vacuum, however this approach leads to mechanical excitation \cite{Ayrapetyan99}, \cite{ijms21218033}, and was later discarded.

\begin{figure}[htp]
\centering
\subfigure[]{\includegraphics[width=.49\textwidth]{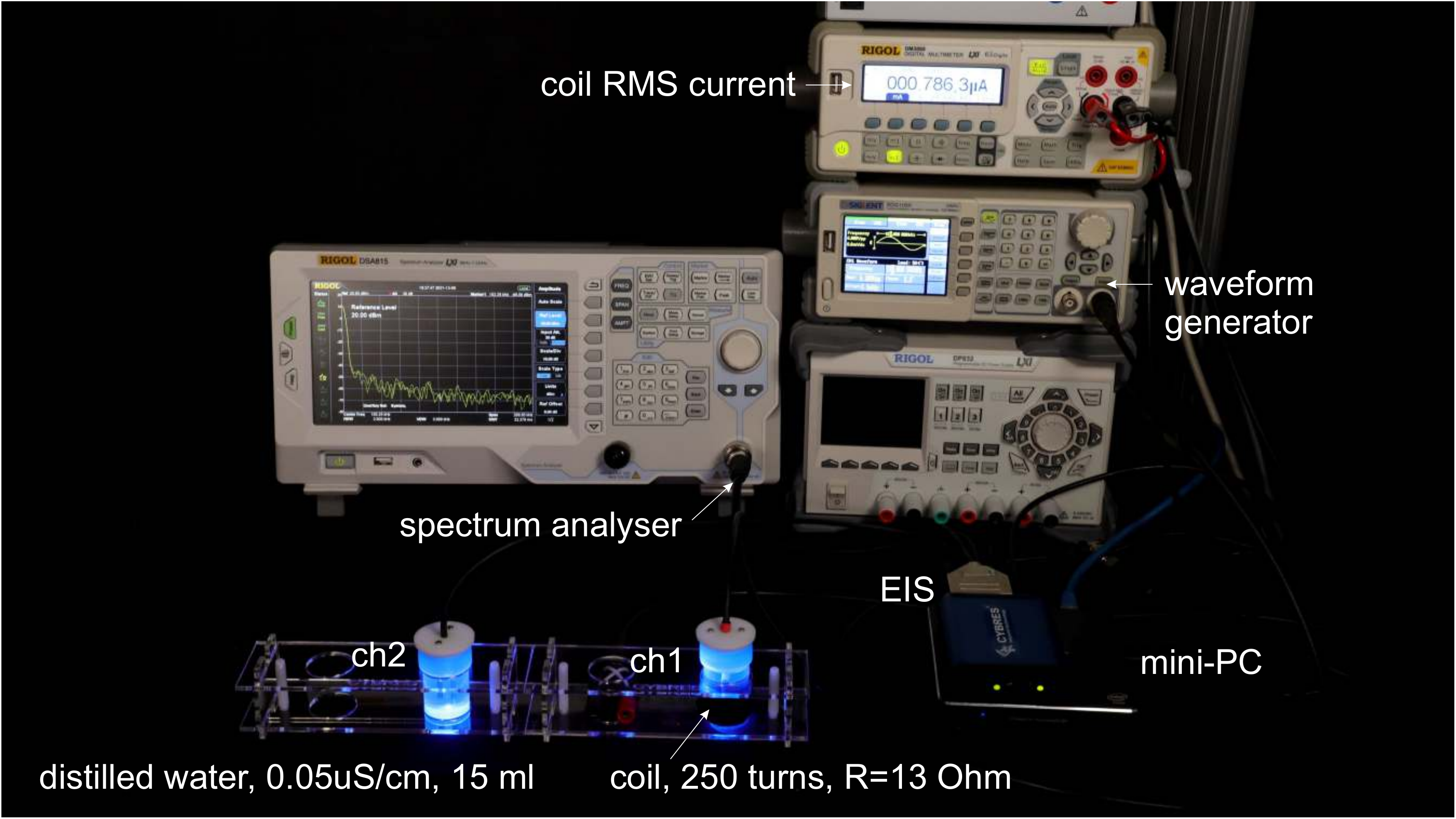}}
\subfigure[]{\includegraphics[width=.49\textwidth]{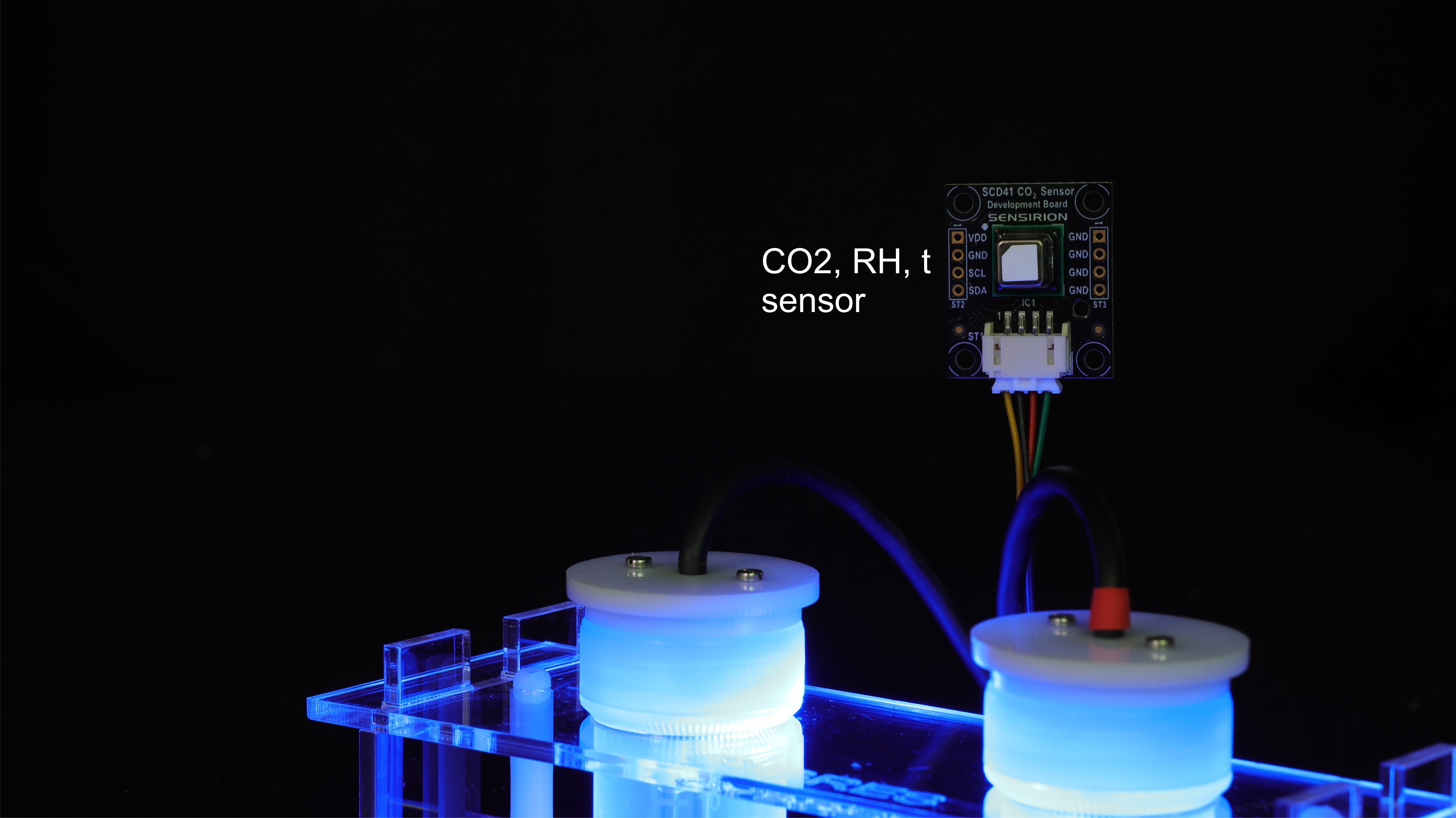}}
\caption{\small \textbf{(a)} Image of the setup (without thermo-insulating boxes and \ce{CO_2} equipment); \textbf{(b)} \ce{CO_2}, RH, t sensor, installed close to measurement containers. EIS excitation by 490nm light is on for demonstration purposes (experiments have been performed in darkness).
\label{fig:setup}}
\end{figure}

Electrochemical measurements \cite{Kernbach17water} are performed by differential impedance spectrometer in continuous and fast EIS mode (CYBRES EIS spectrometer \cite{CYBRES_UserManual}), see the setup in Fig. \ref{fig:setup}. One channel represents a control sample without any treatment, another is the experimental one (containers have holes for gas exchange). Since the level of fluids in cells is varied between experiments, e.g. for open surface scenarios, the impedance is shown without the cell constant. Motivated by \cite{Spangenberg21}, the same measurements have been performed with UV optical absorption spectroscopy (ASEQ instruments, deuterium lamp) and compared with EIS results. Since the low energy excitation $10^{-8}-10^{-9}$J/mL is below the excitation threshold for ROS pathway with chemiluminescence (about 0.15J/mL \cite{Gudkov11}), the single photon detector or avalanche photodetector was not used. Water samples are excited by magnetic field (solenoid on experimental channel) generated by broadband noise. Intensity of magnetic field varies between 78$\mu T$ and 942$\mu T$ (estimated by RMS value of current 5-30 mA, solenoid 4.836 mH, e.g. 5 mA RMS current forms 157.08$\mu T$ magnetic field with stored energy $6.08^{-8}$J or $4.05^{-9}$J/mL).

\section{Measurement results}

\subsection{Control measurements}

\begin{figure}[ht]
\centering
\subfigure[\label{fig:openOpen6}]{\includegraphics[width=.49\textwidth]{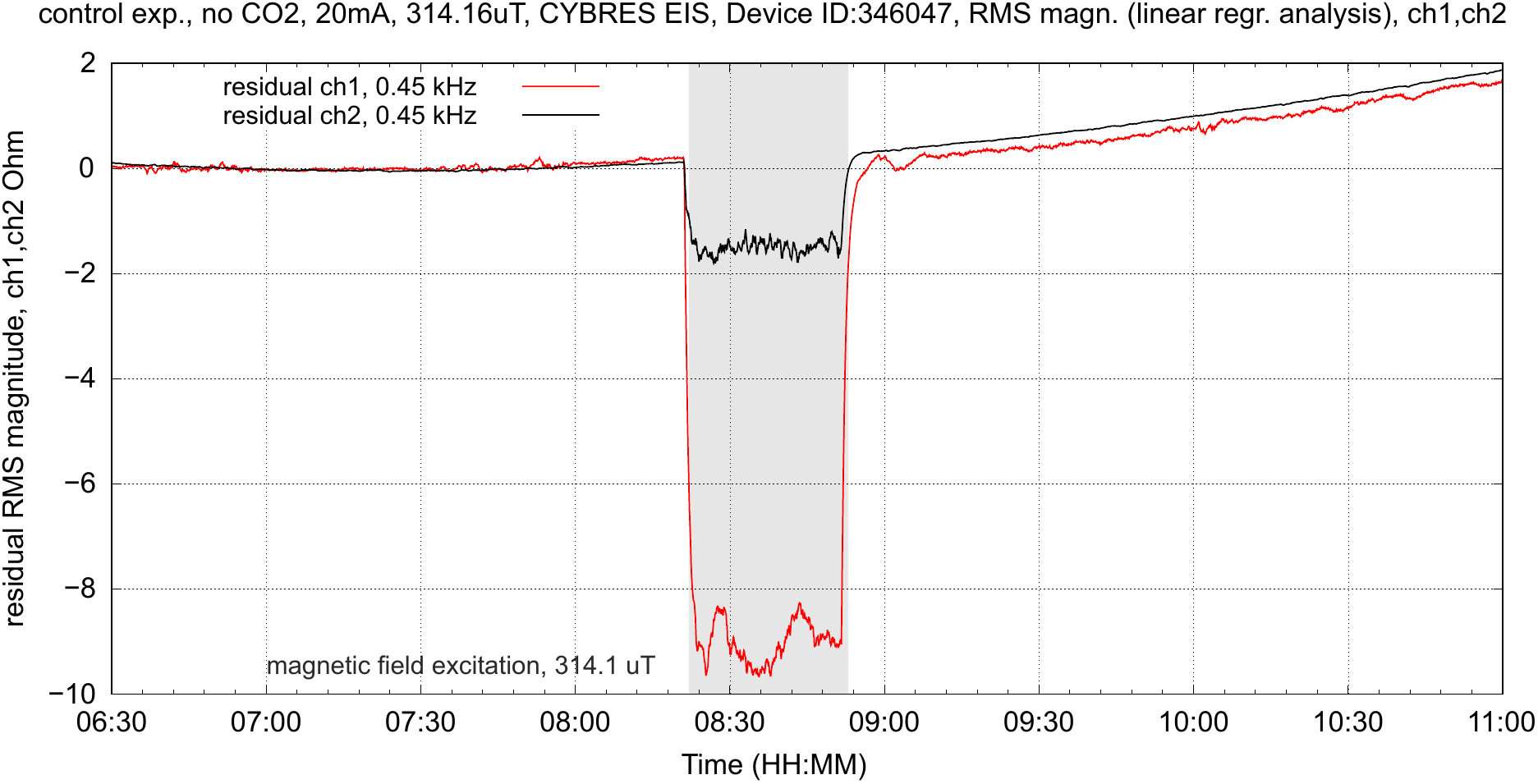}}
\subfigure[\label{fig:openOpen2}]{\includegraphics[width=.49\textwidth]{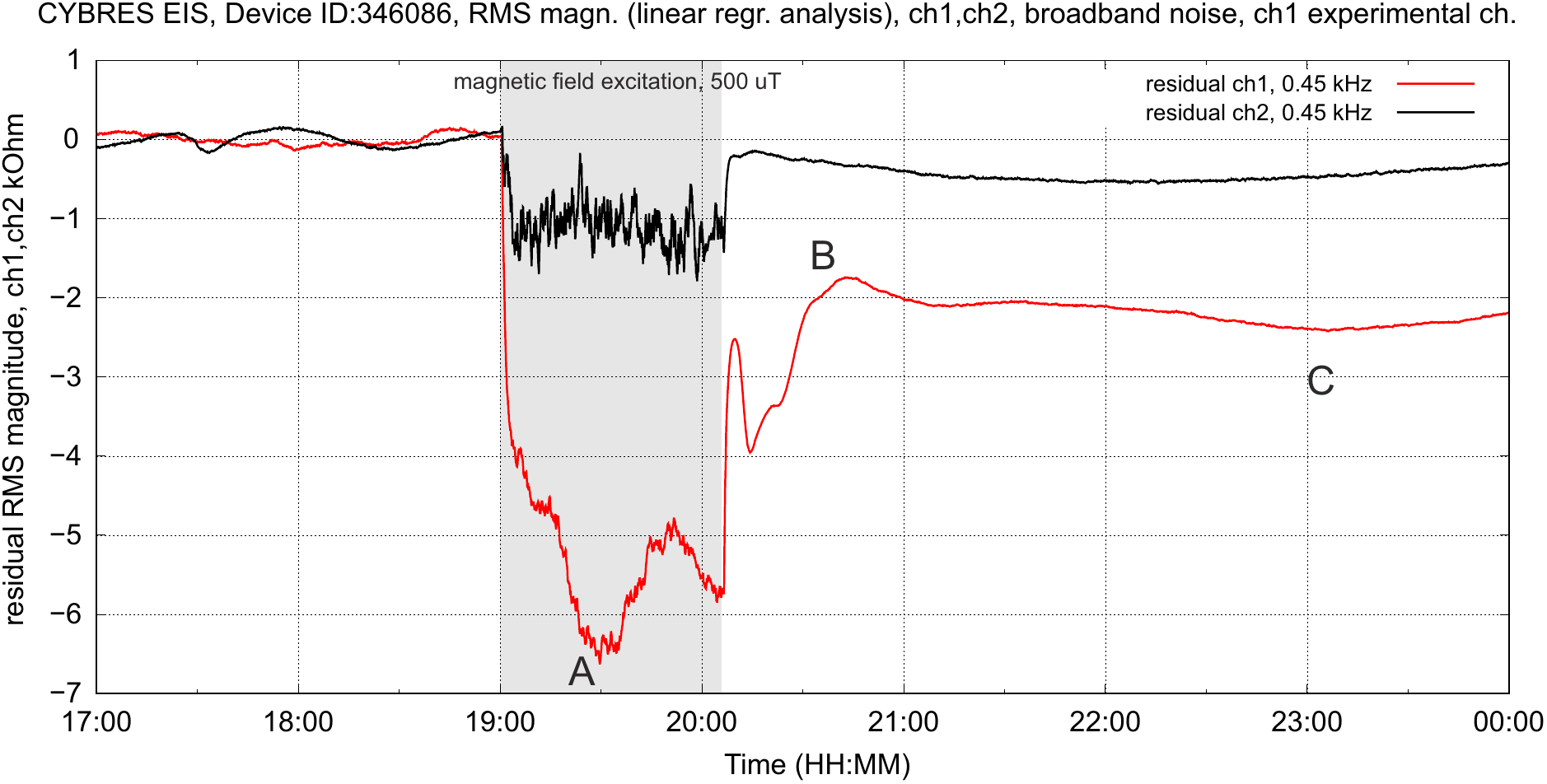}}
\subfigure[\label{fig:h2o2}]{\includegraphics[width=.49\textwidth]{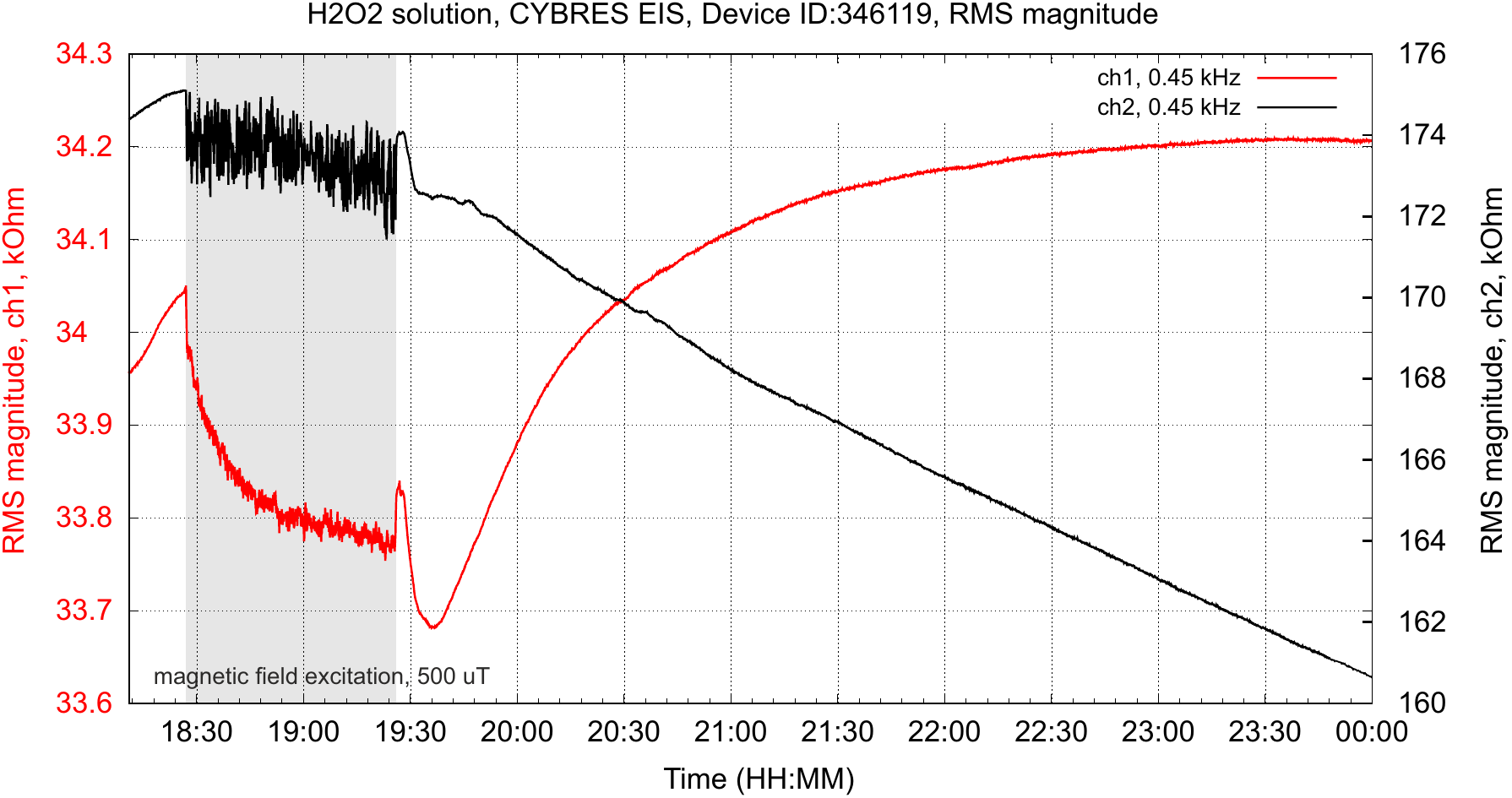}}
\caption{\small The 'low-\ce{CO_2} scenario', measurements of impedances: \textbf{(a)} Control measurements without access to \ce{CO_2} at 314.16$\mu T$ with 20mA RMS current; {\textbf{(b)}} Experiment with access to \ce{CO_2} in open-air conditions, about 500$\mu T$ with handling of samples; \textbf{(c)} Experiments with adding about 0.03ml 3\% solution \ce{H_2O_2} to 15 ml water, 700 $\mu T$, open-air conditions. See supplementary Fig. \ref{fig:controlImpDynamicsSup} for temperature and \ce{CO_2} dynamics in these experiments.
\label{fig:controlImpDynamics}}
\end{figure}

\begin{figure}[ht]
\centering
\subfigure[\label{fig:fastEIS1}]{\includegraphics[width=.49\textwidth]{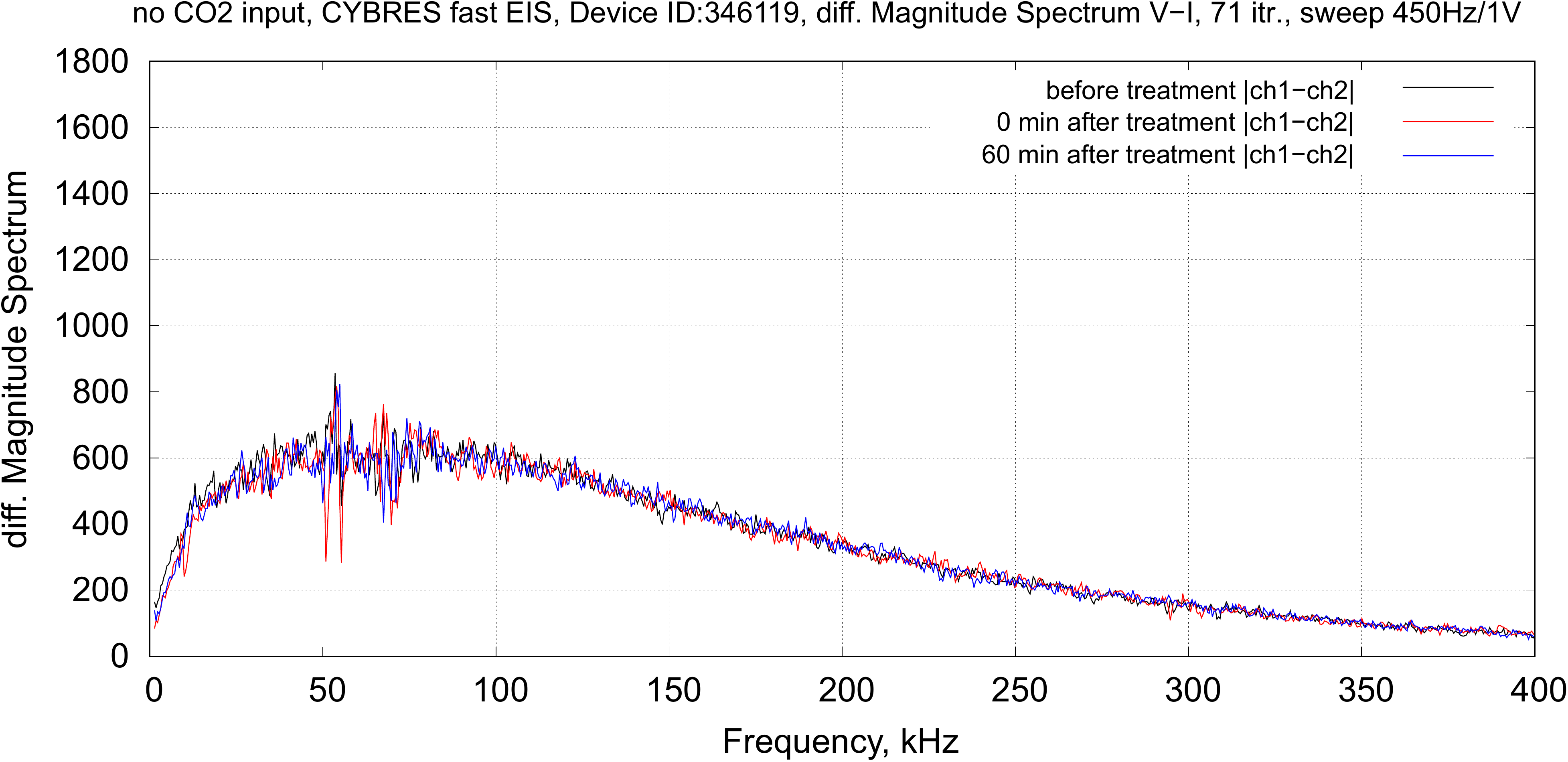}}
\subfigure[\label{fig:fastEIS2}]{\includegraphics[width=.49\textwidth]{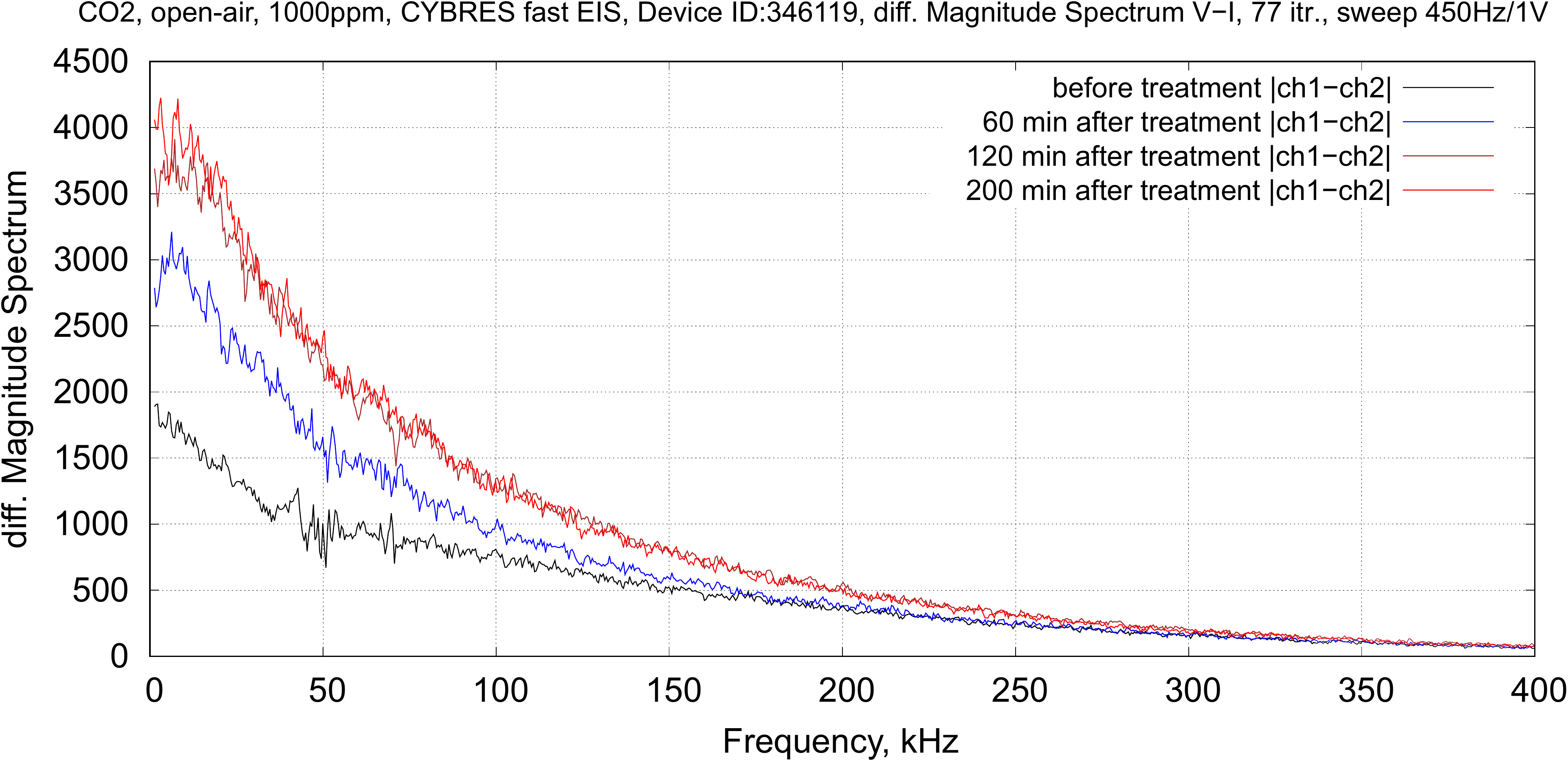}}
\caption{\small The 'open-\ce{CO_2} scenario', differential spectra (difference between control and experimental channels) of fast EIS before or after treatment (broadband noise, field intensity about 500$\mu T$); \textbf{(a)} Control measurements without access to \ce{CO_2} atmosphere; \textbf{(b)} Experiment with access to \ce{CO_2} atmosphere in open-air conditions at 1000ppm (containers are open during treatment and between measurements).
\label{fig:fastEIS}}
\end{figure}

\begin{figure}[htp]
\centering
\subfigure[\label{fig:controlImpDynamics1}]{\includegraphics[width=.49\textwidth]{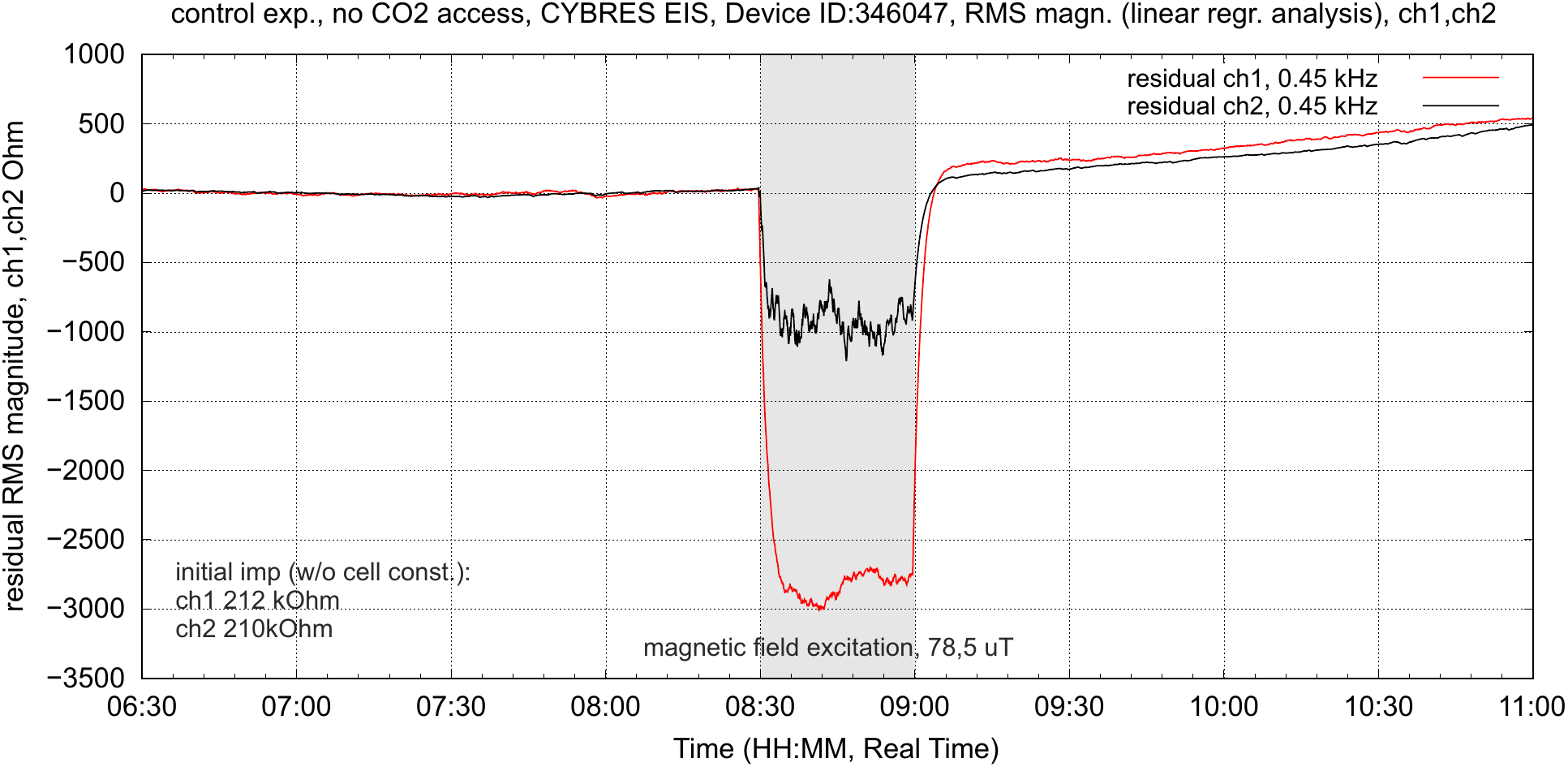}}
\subfigure[\label{fig:controlImpDynamics2}]{\includegraphics[width=.49\textwidth]{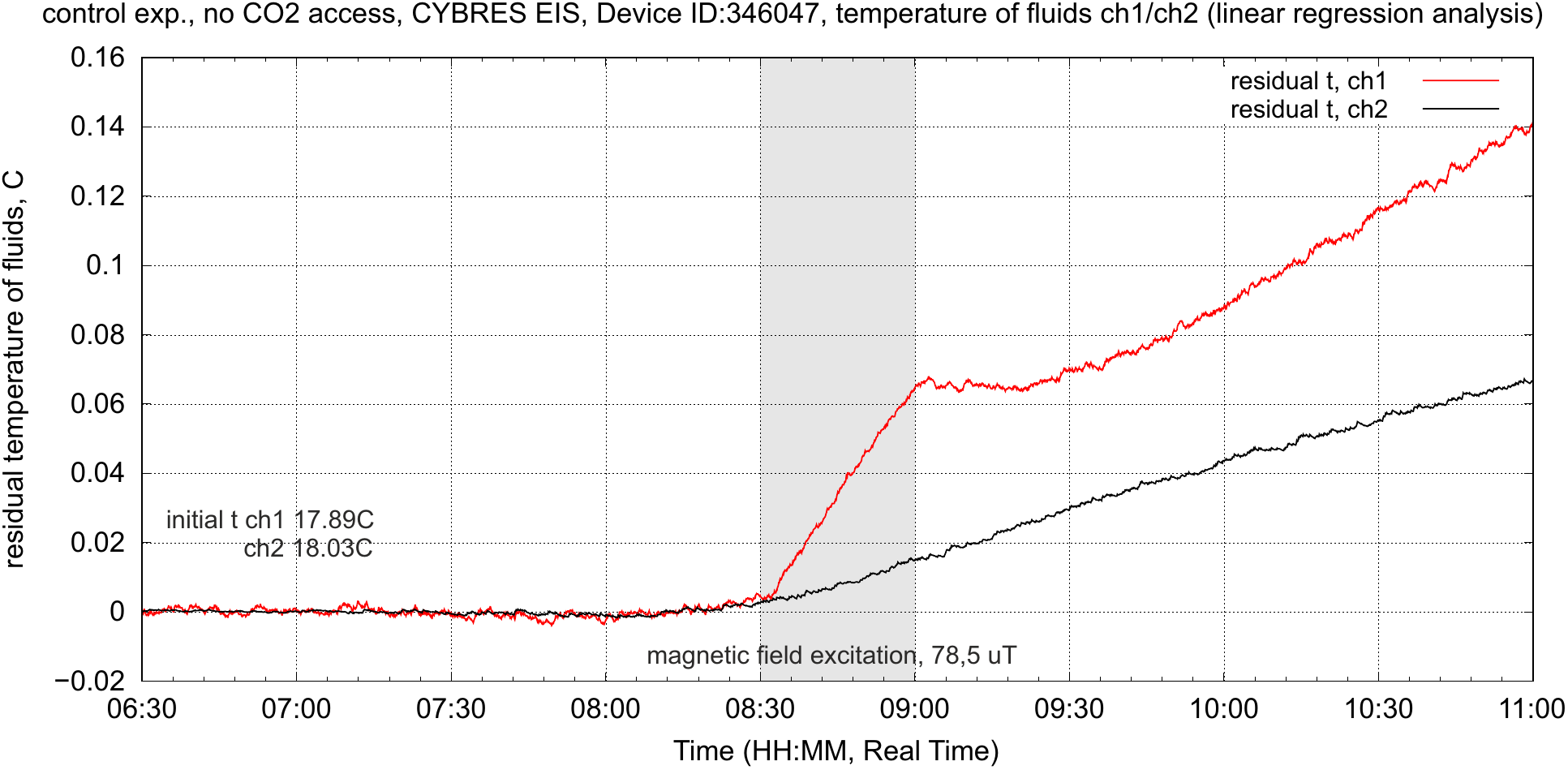}}
\subfigure[\label{fig:closedOpenSupl3}]{\includegraphics[width=.49\textwidth]{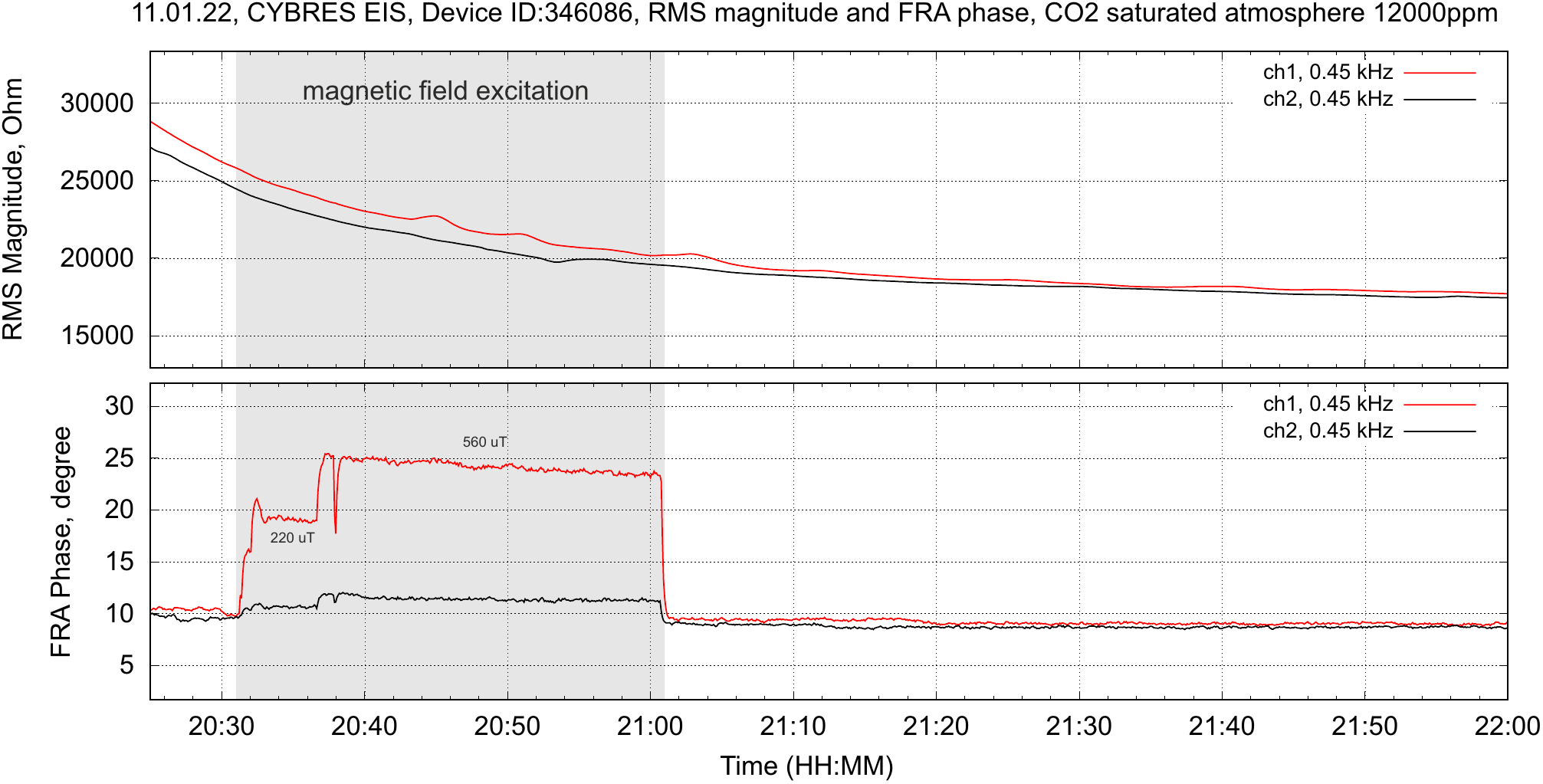}}
\caption{\small Measurements with linear regression of impedances and temperature: \textbf{(a)} Control measurements without access to \ce{CO_2} at 78.5$\mu T$ with 5mA RMS current and \textbf{(b)} dynamics of temperature in this experiment; \textbf{(c)} Phase and magnitude of impedance in \ce{CO_2} saturated phase with magnetic field excitation 220$\mu T$ and 560$\mu T$.
\label{fig:controlImpDynamicsSupl}}
\end{figure}

Applying excitations about 70-700$\mu T$ with closed measurement containers without access to \ce{CO_2}, no essential ionic changes after treatment are observed beyond temperature induced-variations (\ref{eq:t}), see Figs. \ref{fig:openOpen6}, \ref{fig:controlImpDynamics1}. Dynamics of electrochemical impedances in both channels closely follows each other without nonlinear effects. Fast EIS spectra demonstrate also no differences between samples before and after treatments, see Fig. \ref{fig:fastEIS1}.

\begin{figure}[ht]
\centering
\subfigure[\label{fig:fastCO20}]{\includegraphics[width=0.49\textwidth]{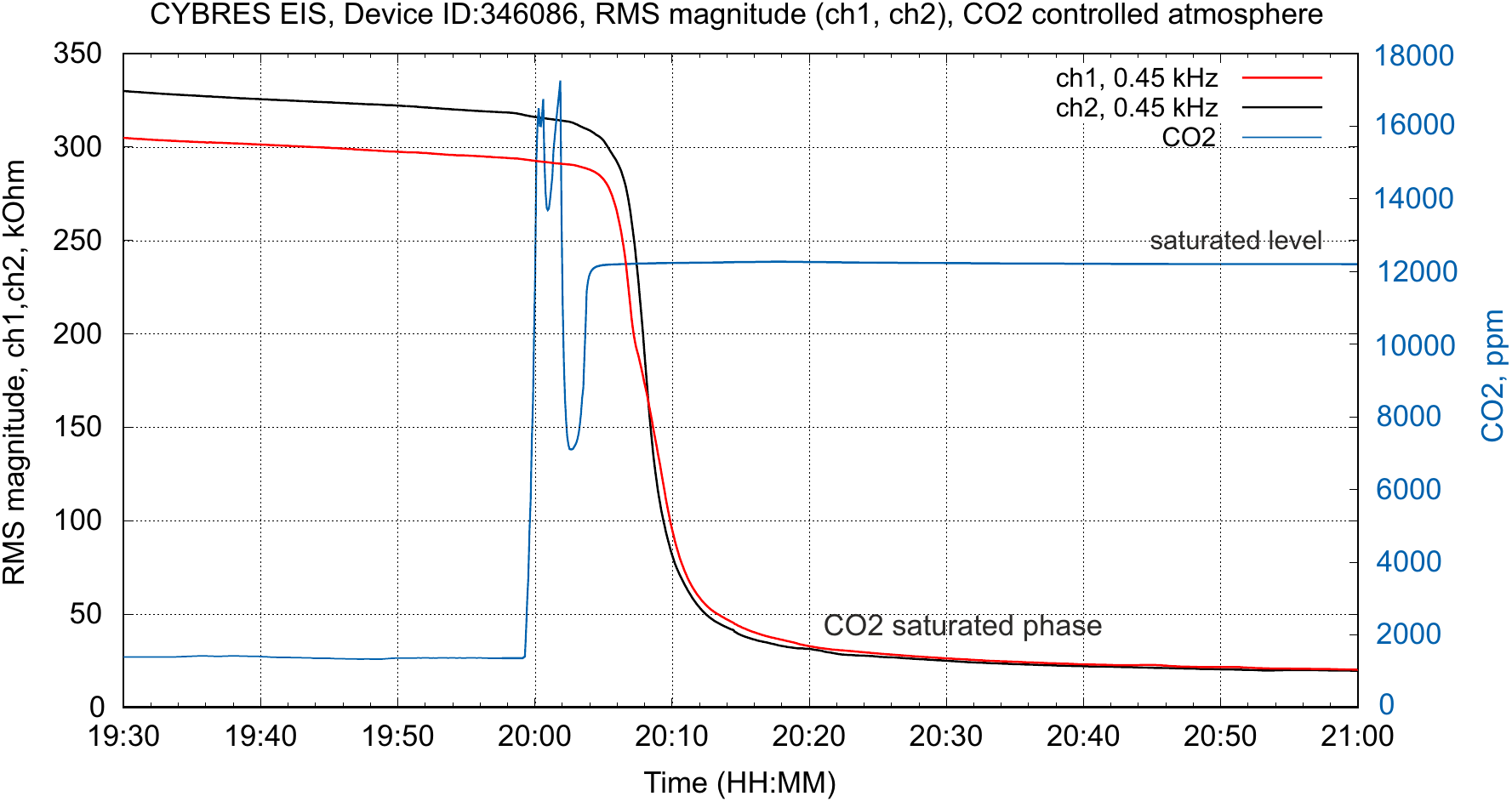}}
\subfigure[\label{fig:fastCO22}]{\includegraphics[width=.49\textwidth]{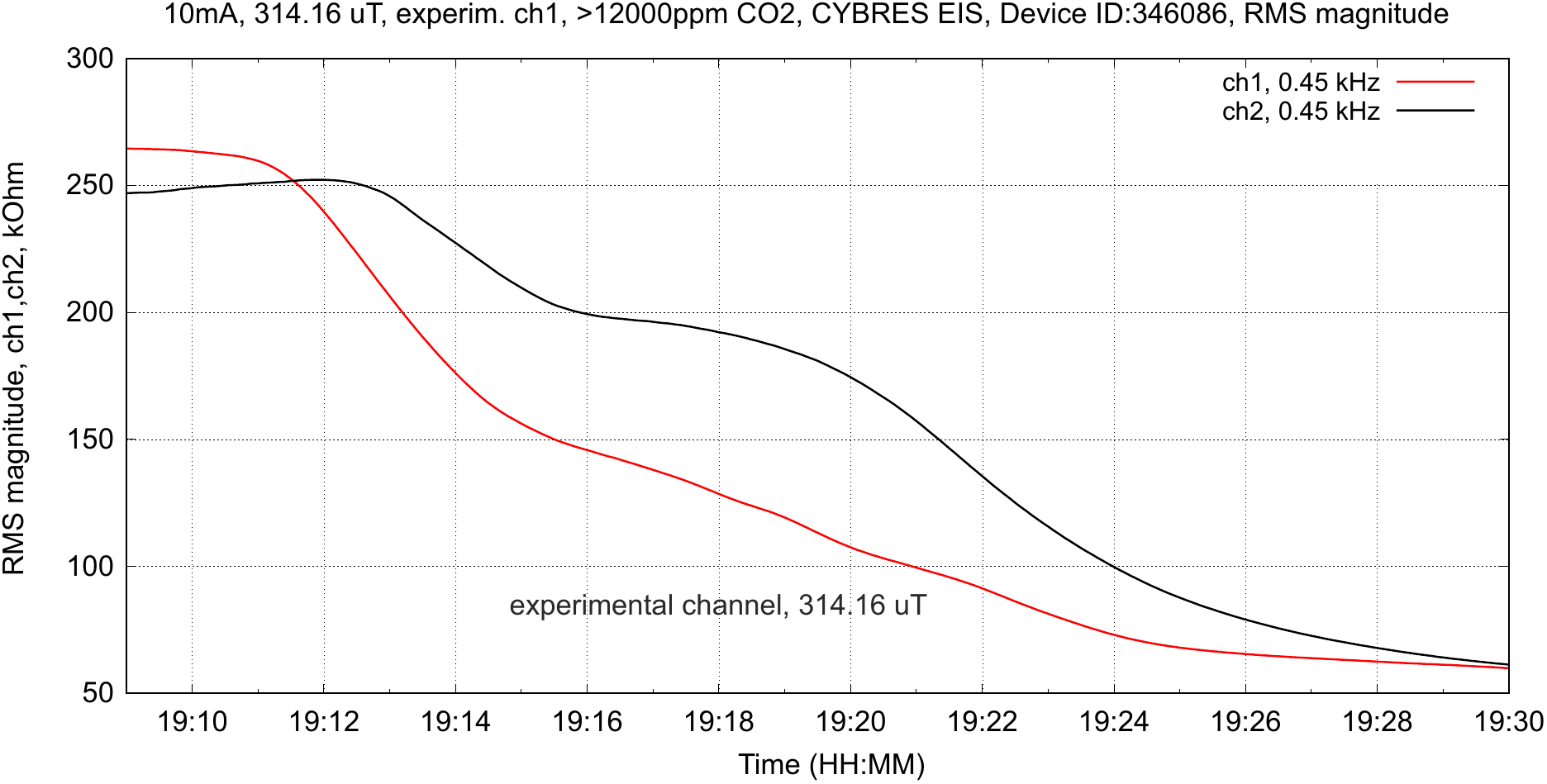}}
\caption{\small The 'fast-\ce{CO_2} scenario', electrochemical impedance (without cell constant), \ce{CO_2} atmosphere is generated in thermo-insulating container 0.0125$m^3$: {\textbf{(a)}} Control attempt without excitation, dynamics of both channels follows each other until ionic saturation is occurred; {\textbf{(b)}} Experiment with excitation by 314.16$\mu T$, samples are inserted into the high-\ce{CO2} phase. See supplementary Fig. \ref{fig:fastCO2Sup} for temperature of fluids during this experiment.
\label{fig:fastCO2}}
\end{figure}

The experiments with \ce{CO_2} saturation are shown in Fig. \ref{fig:fastCO20}, where a fast degradation of impedance within 10-15 min up to the saturation level is observed. Exposure by magnetic field in the \ce{CO_2}-saturated phase does not demonstrate any significant effects in EIS dynamics, see Fig. \ref{fig:closedOpenSupl3}. Control attempts with \ce{H_2O_2} solutions in closed containers have another dynamics, the electrochemical markers are always present, but their intensity and slope (ionic reactivity) depend on \ce{CO_2} input, see Sec. \ref{sec:elecrochemicalMarkers}.

\subsection{Measurements with \ce{CO_2} input}

\begin{figure}[ht]
\centering
\subfigure[\label{fig:fastCO21}]{\includegraphics[width=.49\textwidth]{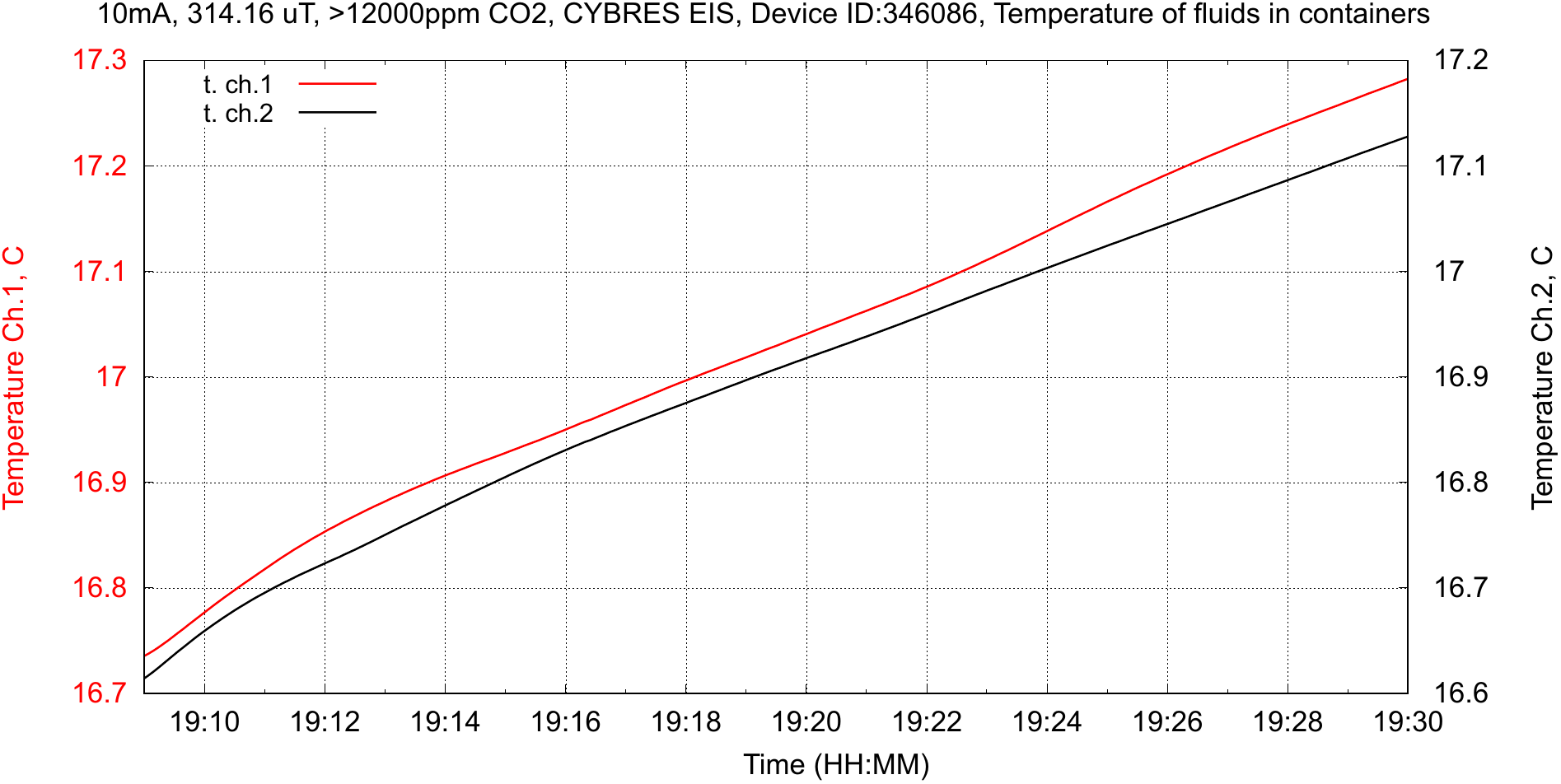}}
\caption{\small Supplementary image to Fig. \ref{fig:fastCO2}: the 'fast-\ce{CO_2} scenario', electrochemical impedance (without cell constant) and temperature. {\textbf{(a)}} temperature of fluids during the experiment with excitation by 314.16$\mu T$, samples are inserted into the high-\ce{CO2} phase, temperature fluctuations due to heating by the solenoid are negligible compared to its own dynamics.
\label{fig:fastCO2Sup}}
\end{figure}

These experiments have been performed in four different scenarios, depending on temperature dynamics, surface tension and amount of \ce{CO_2}:

\textbf{1)} Most evident results are obtained in the 'fast-\ce{CO_2} scenario' with a high concentration of \ce{CO_2}, where the experimental channel demonstrated a faster degradation of impedance (higher ionic reactivity in excited channel), see example in Fig. \ref{fig:fastCO21}. For randomization, all trials are grouped in series of 10 attempts, where positions in the thermobox, containers with holes for gas, size of holes and timing of \ce{CO_2} input are varied.

\textbf{2)} Samples in the 'slow-\ce{CO_2} scenario' are exposed either in \ce{CO_2}-controlled atmosphere with 2000-5000ppm or in a normal atmosphere with 500-1200ppm. Here three different phases, denoted as A, B, and C, can be distinguished, see Fig. \ref{fig:openOpen2} and discussion about electrochemical markers in Sec. \ref{sec:elecrochemicalMarkers}. The slow-\ce{CO_2} scenario requires longer exposure times, results may very between attempts depending on \ce{CO_2} and intensity of magnetic fields.

\begin{figure}[ht]
\centering
\subfigure[\label{fig:viscosity1}]{\includegraphics[width=.49\textwidth]{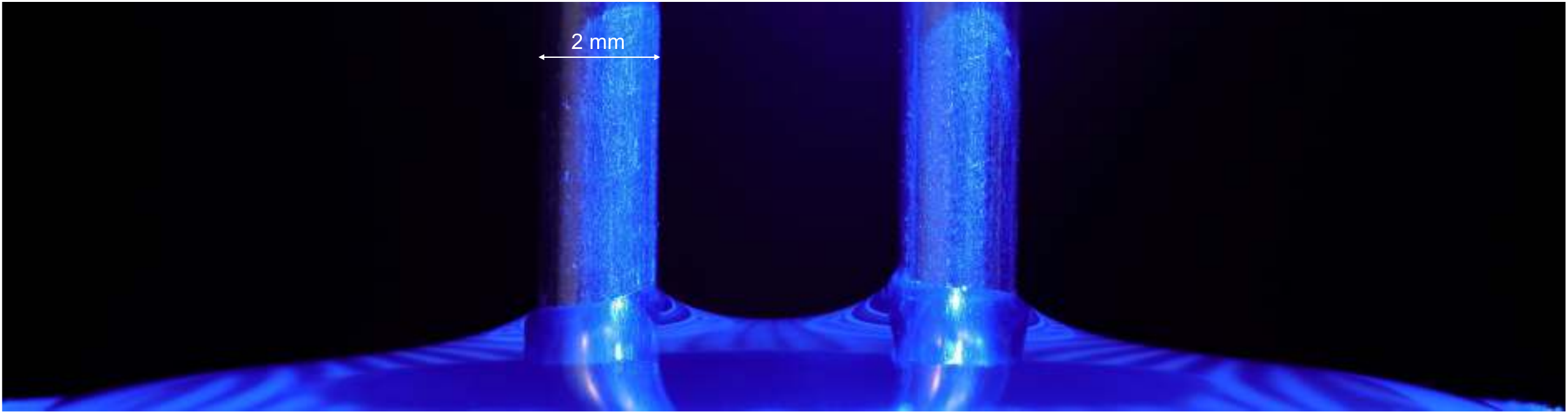}}
\subfigure[\label{fig:viscosity2}]{\includegraphics[width=.49\textwidth]{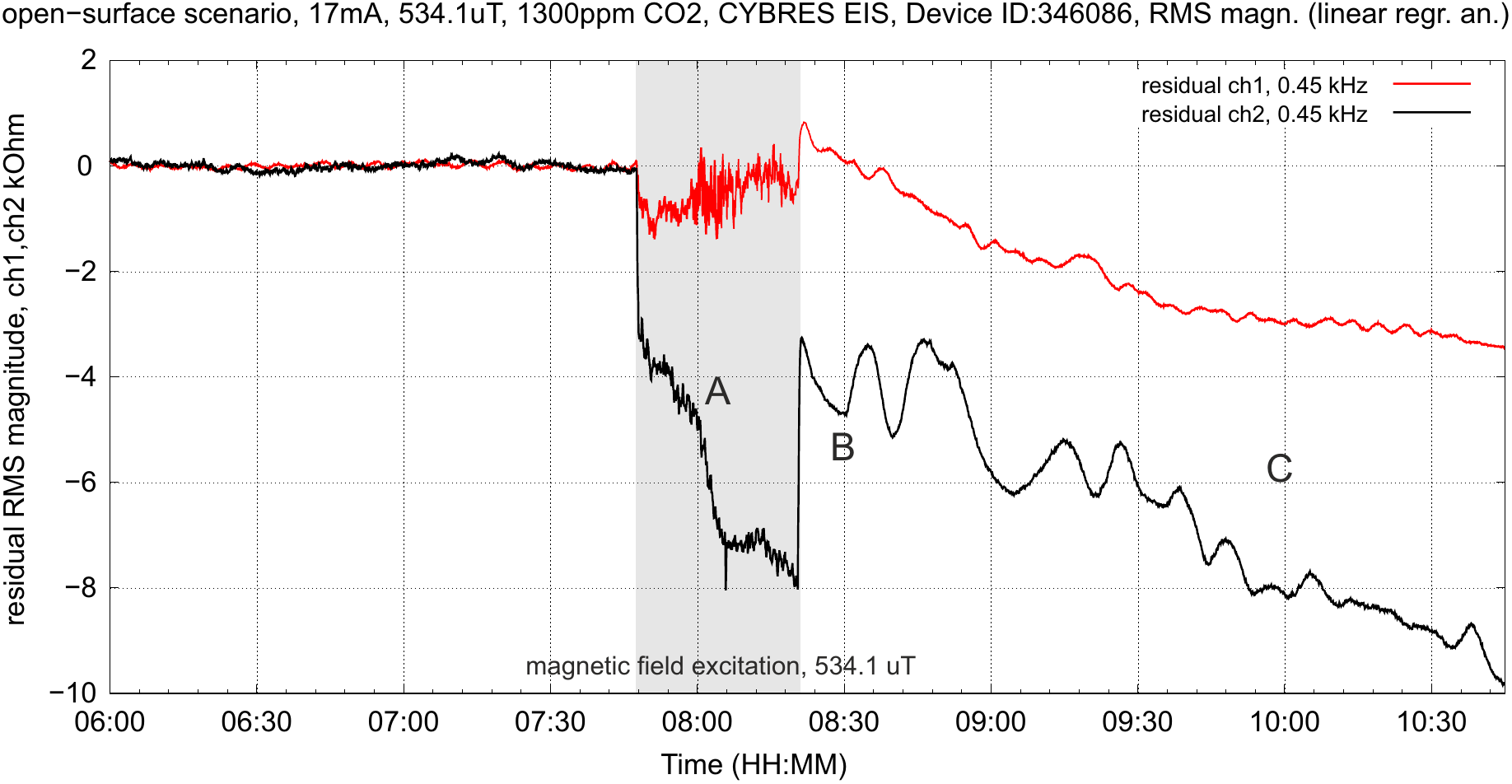}}
\subfigure[\label{fig:viscosity3}]{\includegraphics[width=.49\textwidth]{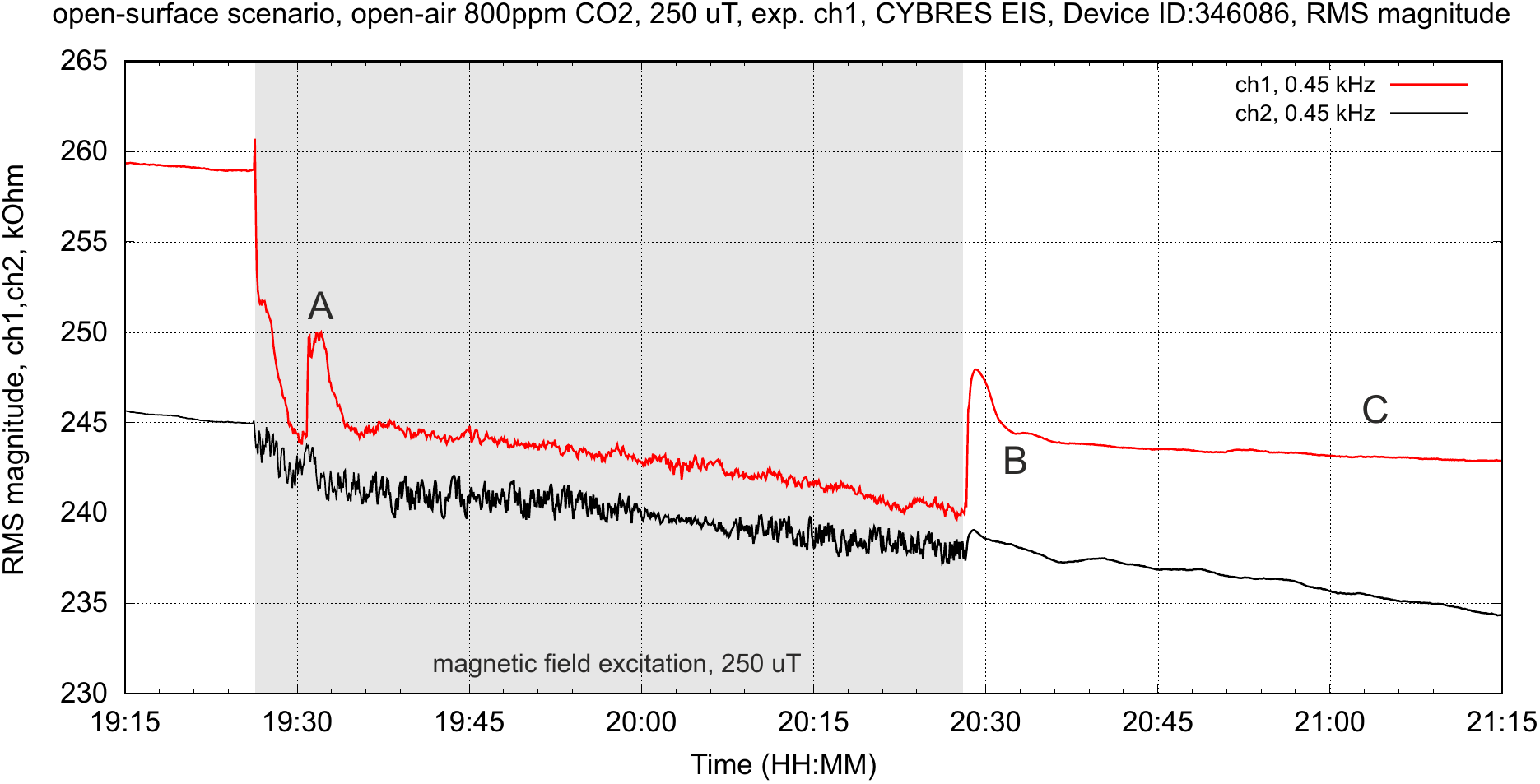}}
\caption{\small The 'open surface scenario': \textbf{(a)} General setup, both channels have the same conditions of open-surface electrodes. Dynamics of electrochemical impedances (without cell constant, no mechanical handling of samples): \textbf{(b)} the channel 2 is exposed to magnetic field at 534.1$\mu T$, 1300ppm \ce{CO_2}, linear regression, appearance of oscillations after excitation is well visible and \textbf{(c)} the channel 1 is exposed to magnetic field at about 250 $\mu T$, 800 ppm \ce{CO_2}. The dynamical markers \textbf{A}, \textbf{B} and \textbf{C} are the same as in other \ce{CO_2} and \ce{O_2} pathways.
\label{fig:viscosity}}
\end{figure}

\textbf{3)} Water containers in the third 'open-\ce{CO_2} scenario' are open in a normal atmosphere during treatment and between measurements by electrochemical and UV absorption spectroscopy. Results demonstrate clear differences between experimental and control samples after treatment in low frequency area, see Fig. \ref{fig:fastEIS}. Experimental observations of 'open-\ce{CO_2} scenario' correlate with phenomenology of other experiments \cite{Konuchov95}, \cite{ijms21218033}, which reported better results in open air conditions, see Sec. \ref{sec:discussion}.

\begin{figure}[ht]
\centering
\subfigure[\label{fig:oscillatingDynamics2}]{\includegraphics[width=.49\textwidth]{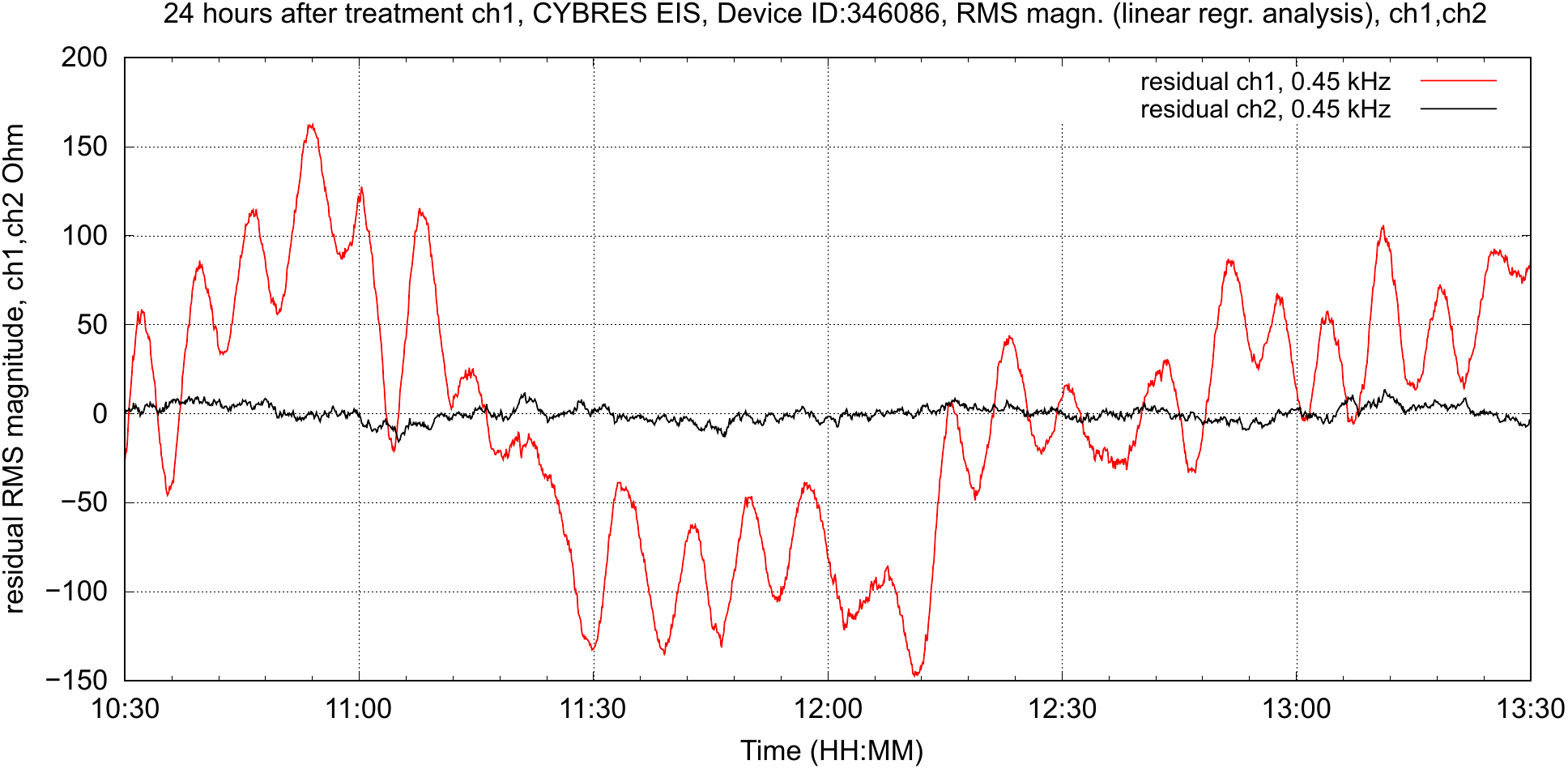}}
\subfigure[\label{fig:oscillatingDynamics3}]{\includegraphics[width=.49\textwidth]{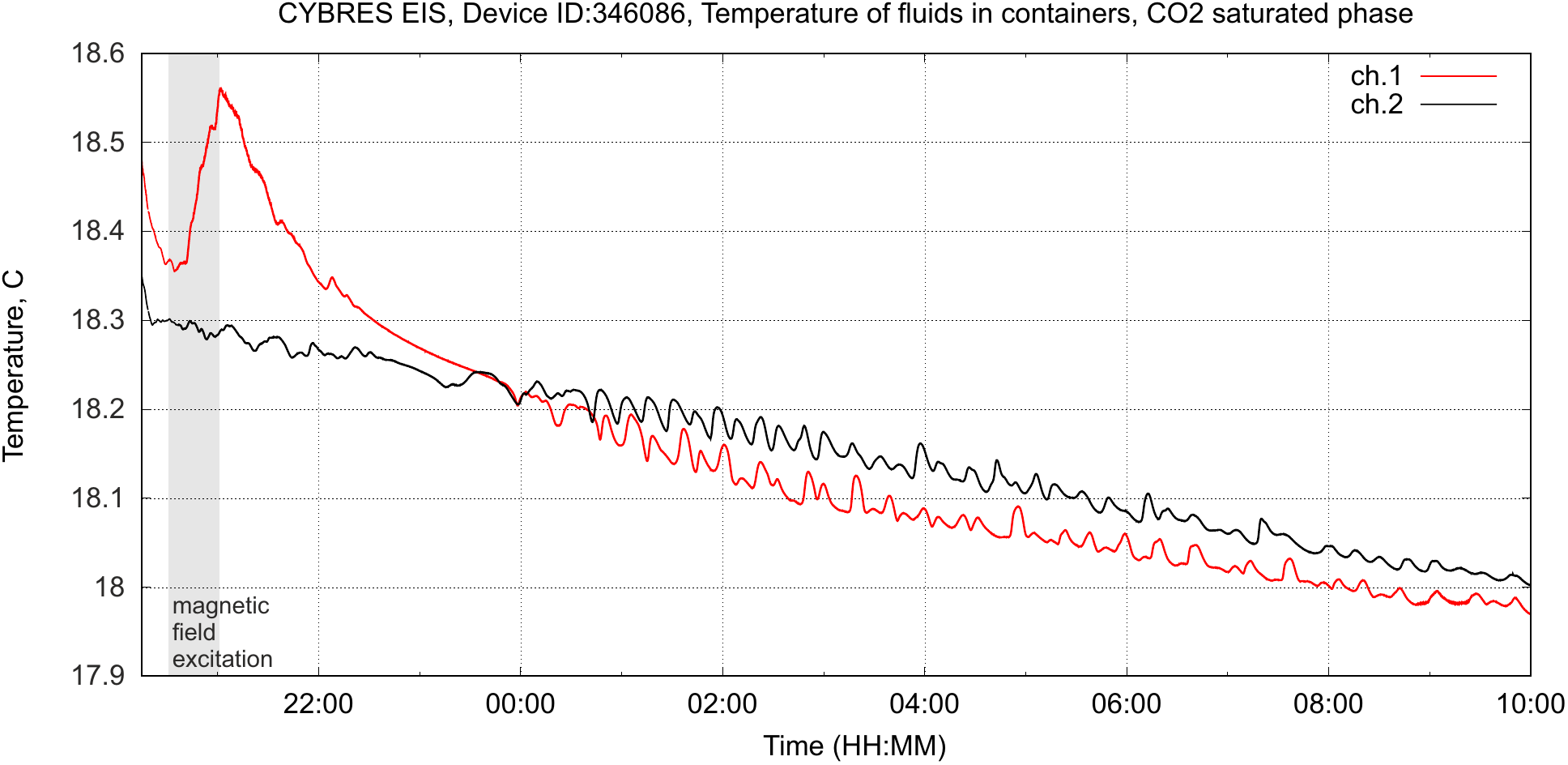}}
\caption{\small Examples of oscillating electrochemical impedances and temperature of fluids after excitation. \textbf{(a)} Electrochemical impedance measurements (without cell constant) in normal atmosphere with 700-1000ppm \ce{CO_2} after treatment (no oscillations in control channel), see supplementary Fig. \ref{fig:oscillatingDynamicsSup}; \textbf{(b)} Appearance of temperature oscillations after excitation in the \ce{CO_2} saturated phase with 80000-14000ppm \ce{CO_2}.
\label{fig:oscillatingDynamics}}
\end{figure}

\textbf{4)} The last 'open surface scenario' is shown in Fig.~\ref{fig:viscosity} and explores the effect of surface tension that according to \cite{Pershin15Biophysics}, \cite{Pershin12Phytosyntesis} should represent independent influencing factor. Electrodes in this experiments are half-removed from fluids so that the surface tension influences the measurement area of electrodes as well as the configuration of electric field under water. Indeed, in this scenario, compared to other cases, the performed measurements showed a higher level of results, even the appearance of oscillations immediately after excitation, see Fig. \ref{fig:oscillatingDynamics}.

Handling control and experimental water samples after excitation (i.e. small mechanical influences imposed on both channels), such as disconnecting cables or preparing spectroscopy, essentially contributes to differentiation of results. Despite handling is similar in both samples, experimental channel with \ce{CO_2} input demonstrates largest variations of impedance after handling (see discussion in Sec. \ref{sec:discussion}) in almost all tested attempts.

\begin{figure}[ht]
\centering
\subfigure[\label{fig:oscillatingDynamics1}]{\includegraphics[width=.49\textwidth]{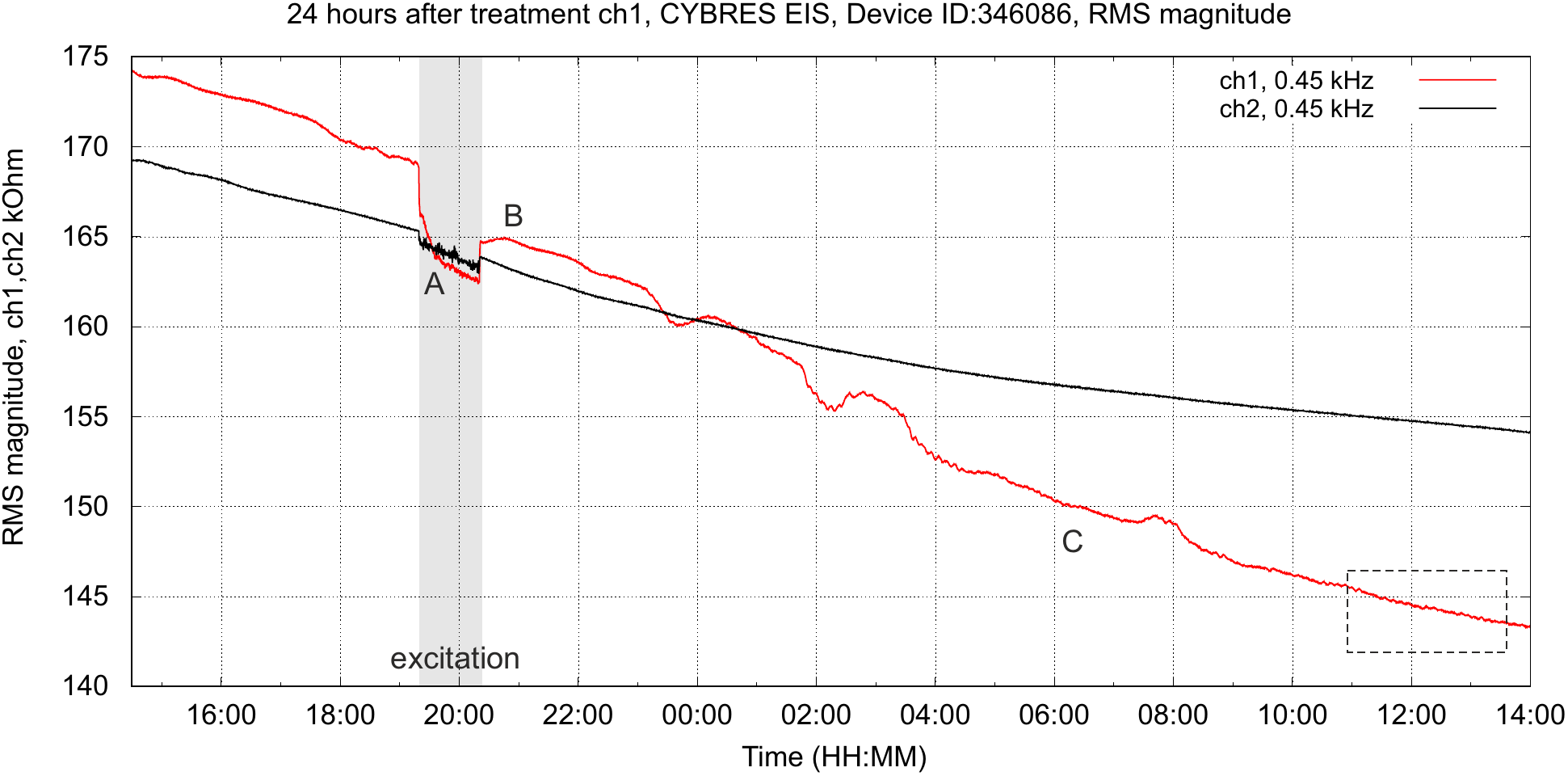}}
\subfigure[\label{fig:oscillatingDynamics4}]{\includegraphics[width=.49\textwidth]{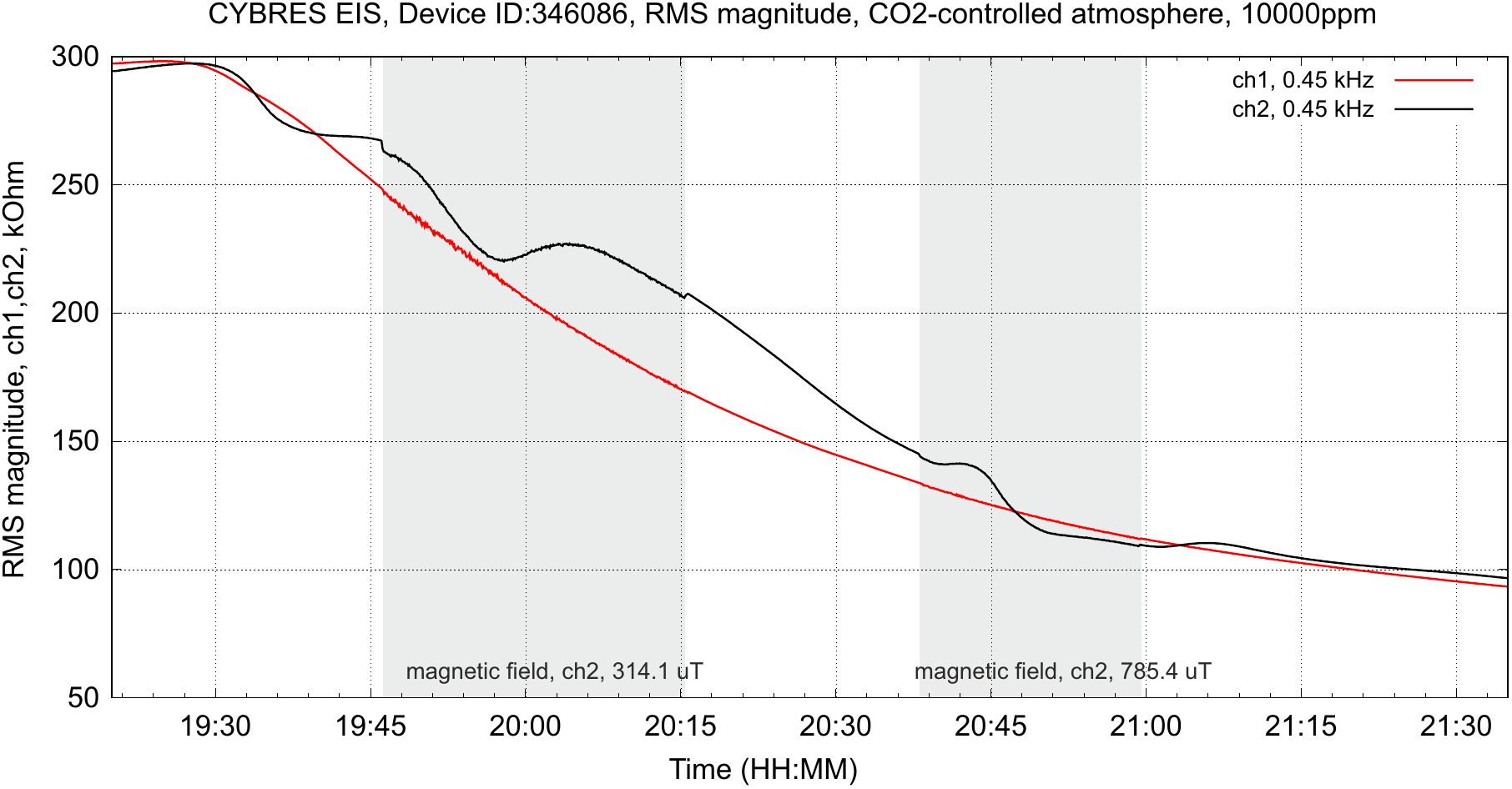}}
\caption{\small Supplementary image to Fig. \ref{fig:oscillatingDynamics}: \textbf{(a)} Electrochemical impedance measurements (without cell constant) in normal atmosphere with 700-1000ppm \ce{CO_2} after treatment, fragment is shown in Fig. \ref{fig:oscillatingDynamics2}; \textbf{(b)} Inducing oscillations of impedance in the fast-\ce{CO_2} scenario by different excitations. 
\label{fig:oscillatingDynamicsSup}}
\end{figure}

Oscillations of electrochemical impedances and temperature of fluids frequently occur in the measured data, see Fig. \ref{fig:oscillatingDynamics}. Typically, they start a few hours after excitation with almost sinusoidal form that slowly transforms to quasi-periodical dynamics with periods between 1 and 60 minutes. Observable duration of oscillations are limited by 24-hours-cycle of measurements. Oscillations appear even in the fast-\ce{CO_2} scenario, where the excitation by magnetic field can stop or induce oscillations, see Fig. \ref{fig:oscillatingDynamics4}.

\subsection{Adding \ce{H_2O_2} solution}

Adding \ce{H_2O_2} (0.03-0.09ml 3\% solution) to 10-15 ml water in experimental channel generates \ce{H_3O^+} and \ce{HO_2^-} ions that essentially decrease impedance from 0.5MOhm up to 30-40kOhm, see Fig \ref{fig:h2o2}. Exciting the solution by weak magnetic field, the impedance first essentially decreases, which points to decreasing of ionic content and indicates possible low-energy transformation mechanisms, see Sec. \ref{sec:discussion} for discussions. After initial phase, the impedance is slowly increasing again that reflects the ionic dissociation (\ref{eq:ionic}) of \ce{H_2O_2}.

\subsection{Electrochemical markers and UV absorption}
\label{sec:elecrochemicalMarkers}

\begin{figure*}[ht]
\centering
\subfigure[\label{fig:global1}]{\includegraphics[width=.49\textwidth]{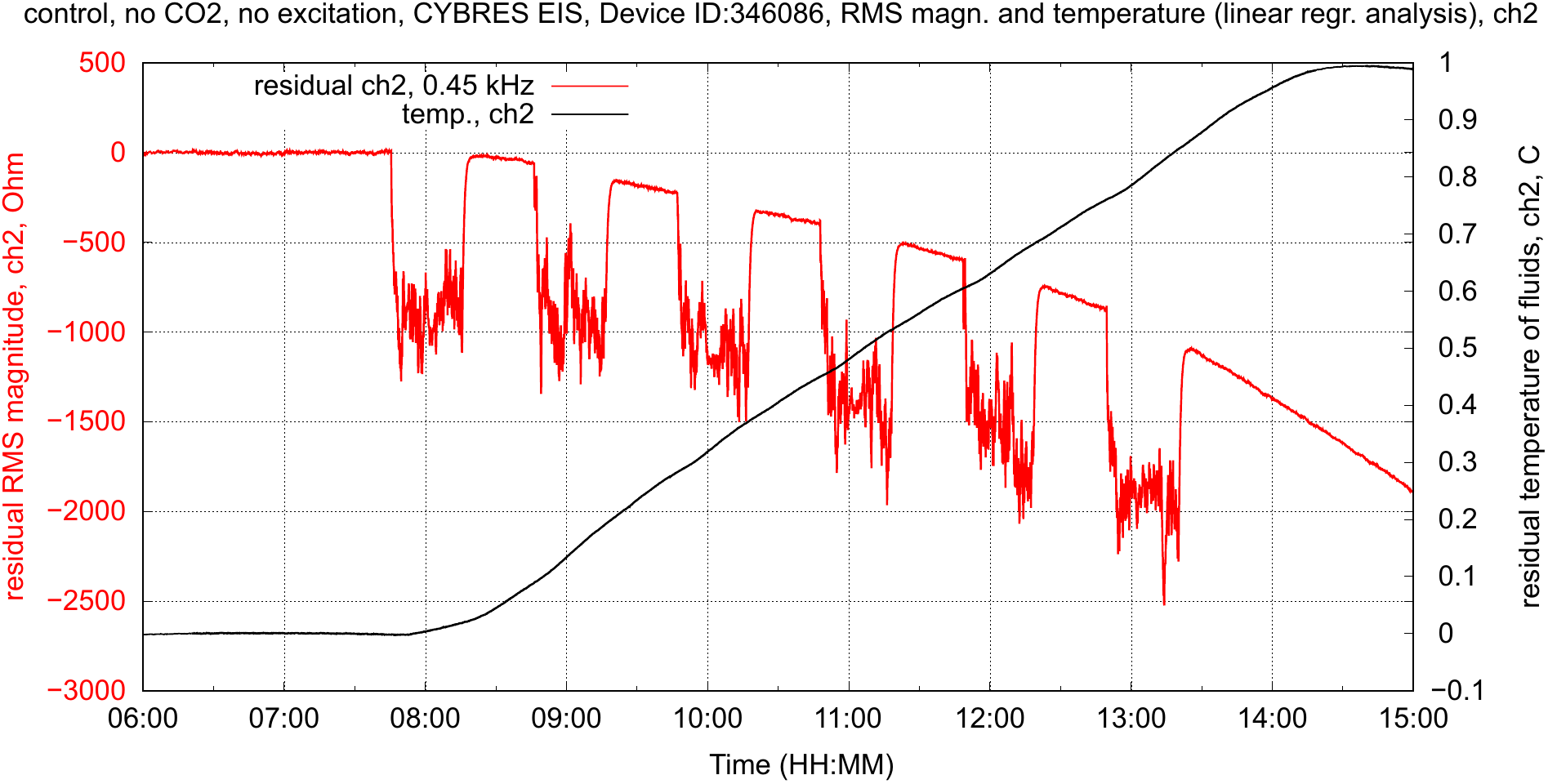}}~
\subfigure[\label{fig:global2}]{\includegraphics[width=.49\textwidth]{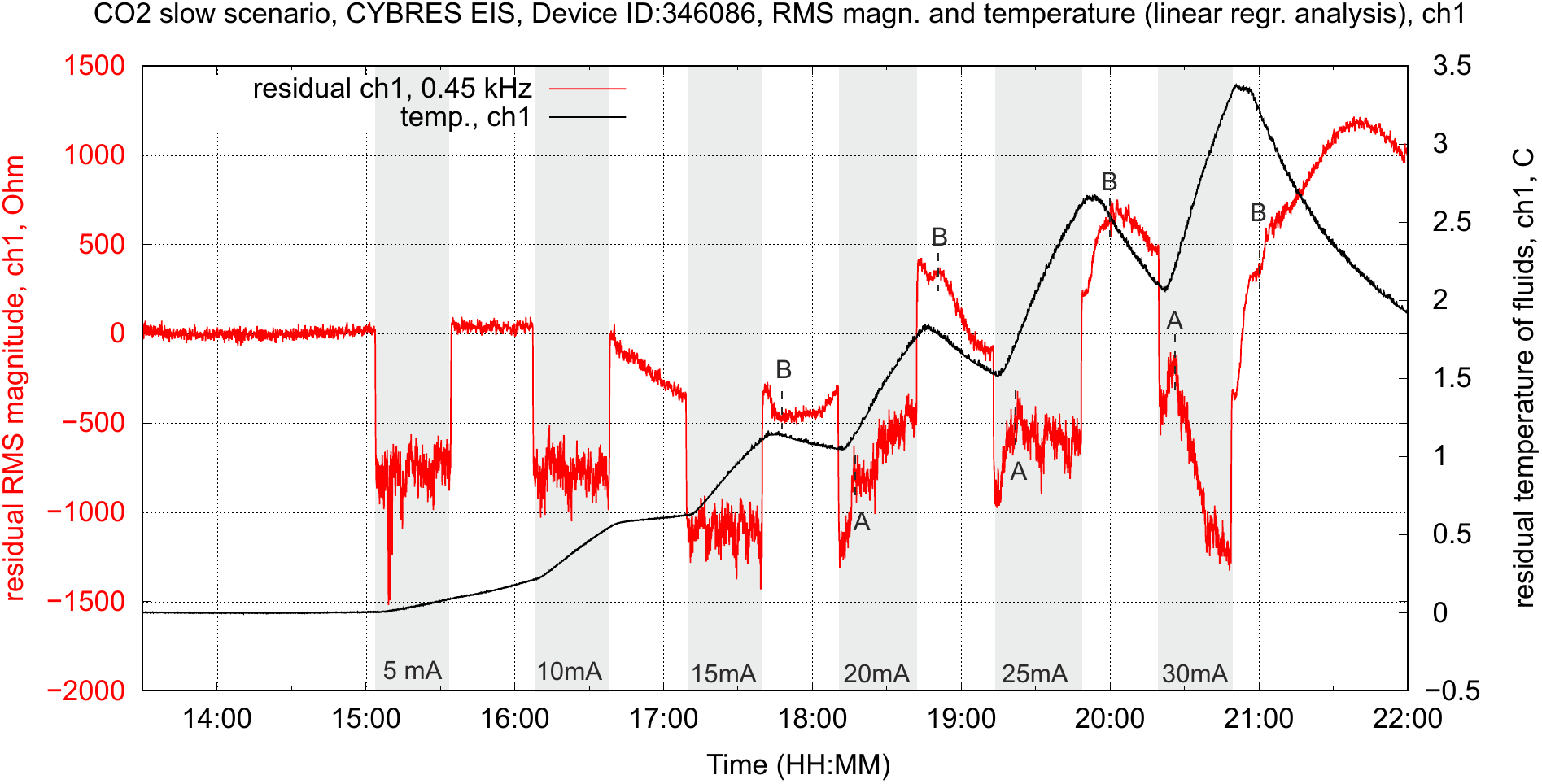}}\\
\subfigure[\label{fig:global3}]{\includegraphics[width=.49\textwidth]{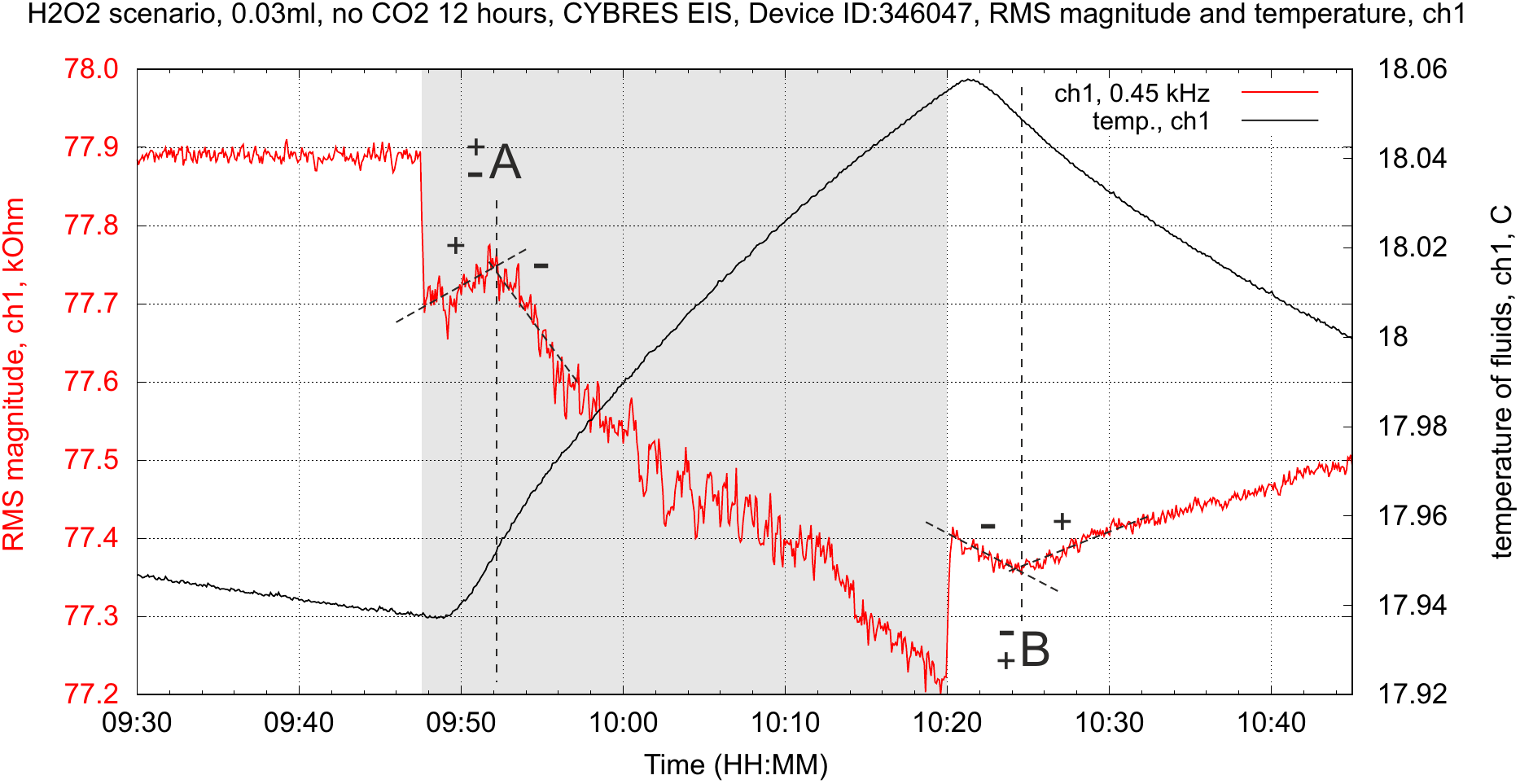}}~
\subfigure[\label{fig:global4}]{\includegraphics[width=.49\textwidth]{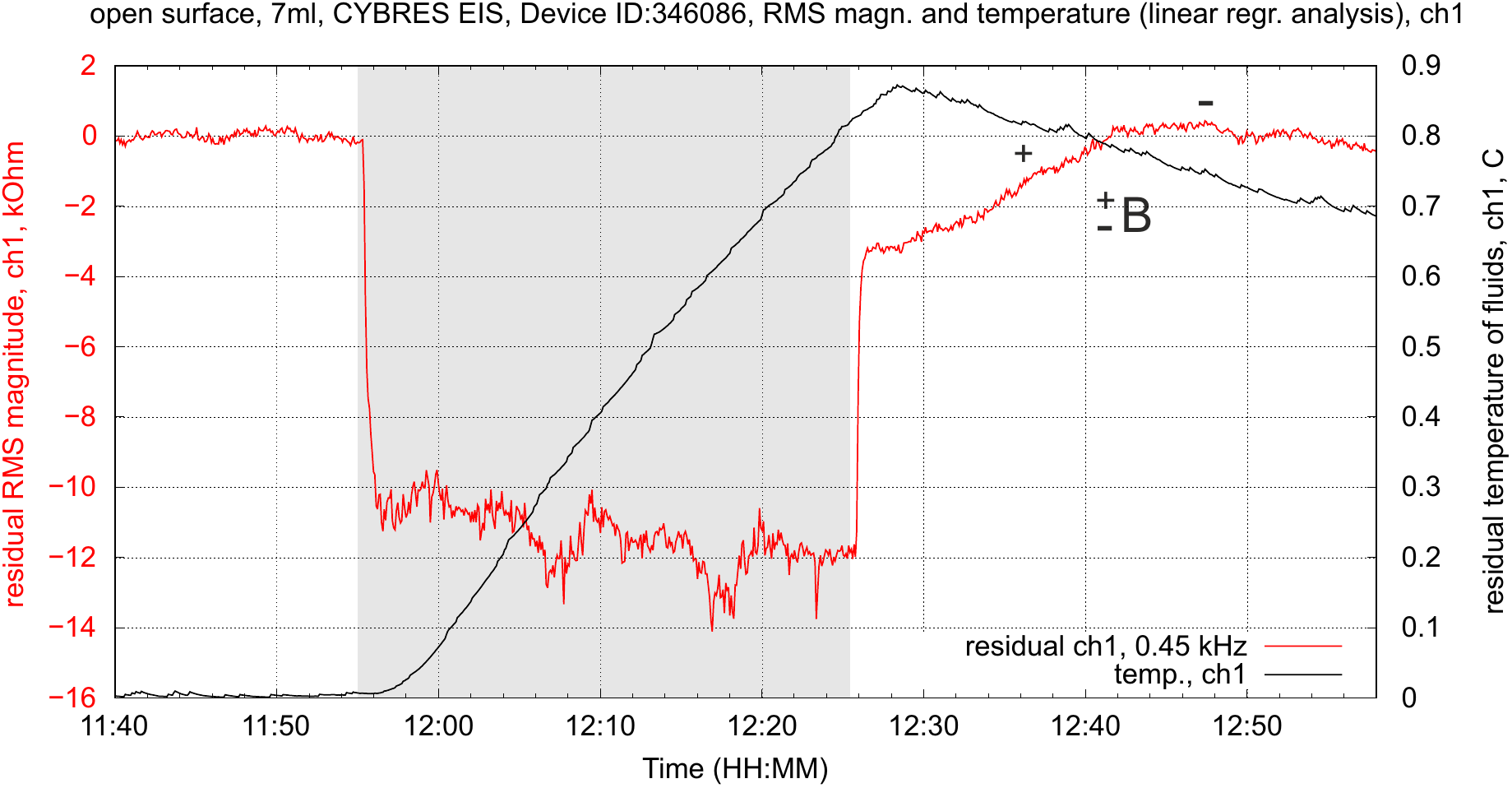}}\\
\subfigure[\label{fig:global5}]{\includegraphics[width=.49\textwidth]{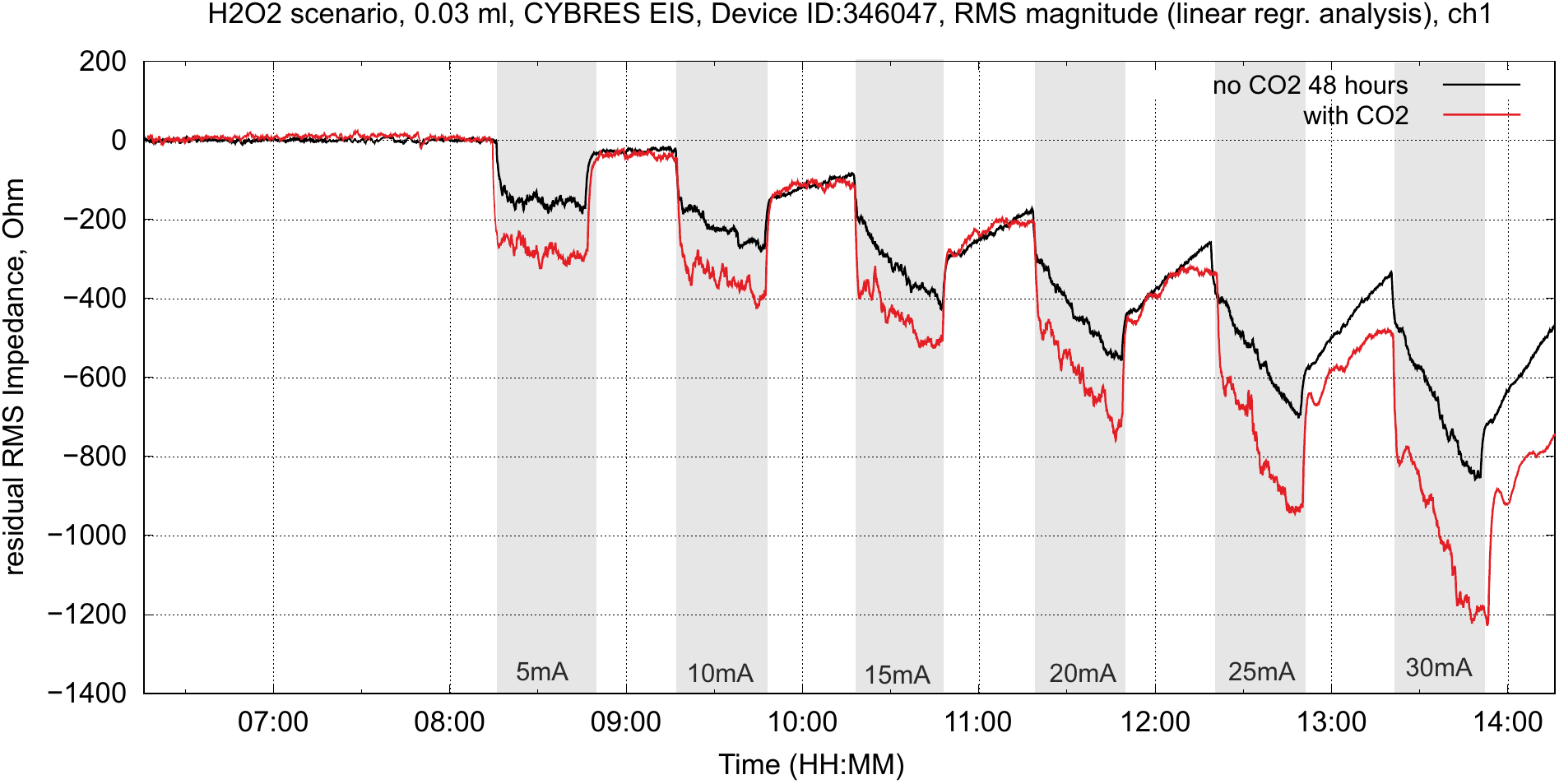}}~
\subfigure[\label{fig:global6}]{\includegraphics[width=.49\textwidth]{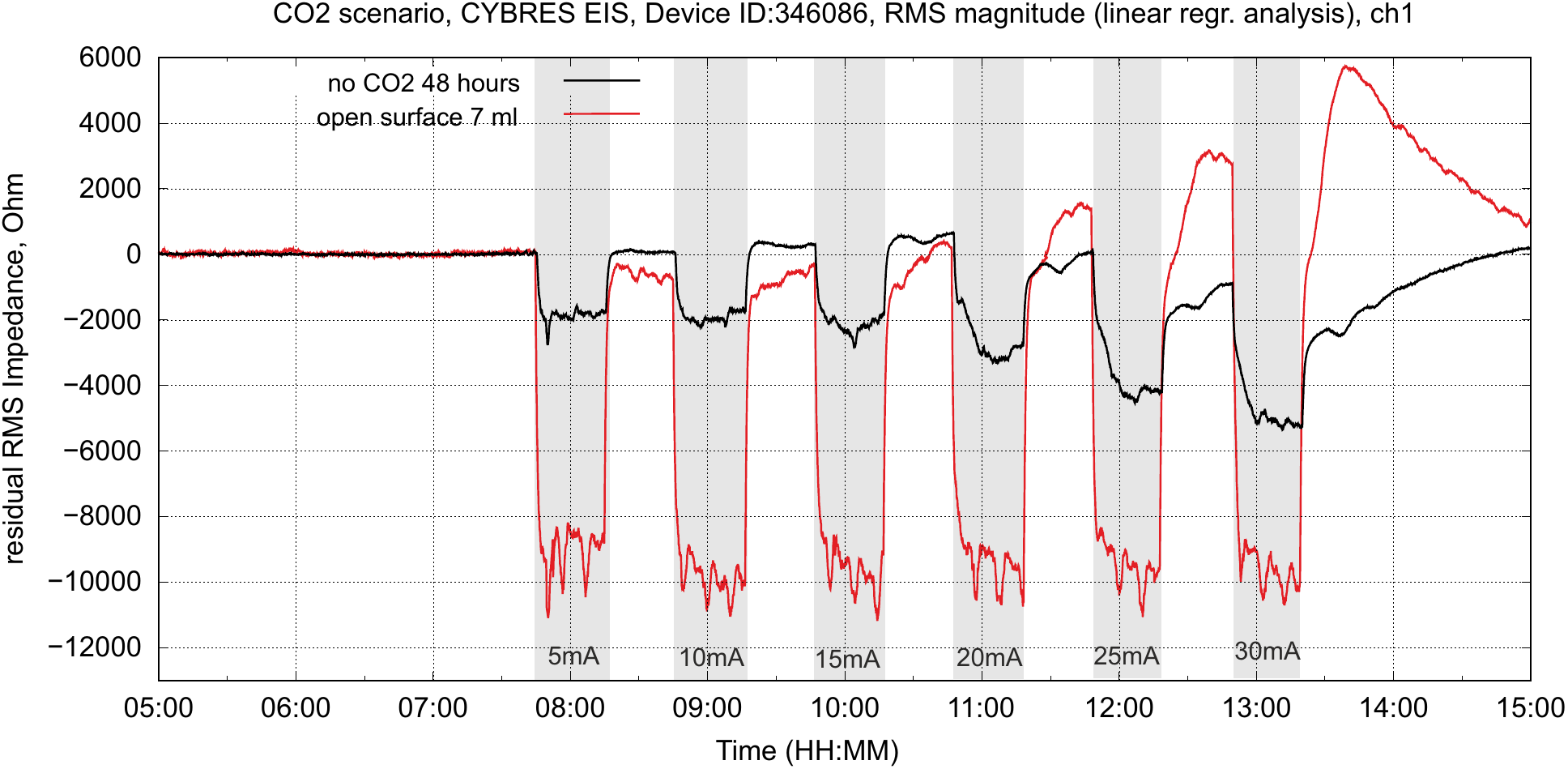}}
\caption{\small Dependency between impedances and temperature of fluids: \textbf{(a)} control channel (no excitations, only EM interferences) with decreasing of impedance; \textbf{(b)} the slow \ce{CO_2} scenario with increasing of impedance triggered by weak excitations. Examples of electrochemical markers: \textbf{(c)} \ce{H_2O_2} scenario; \textbf{(d)} the open surface scenario. Comparison of electrochemical dynamics (ionic reactivity) with/without \ce{CO_2} input and with the same heating impacts: \textbf{(e)} \ce{H_2O_2} scenario and \textbf{(f)} \ce{CO_2}/open surface scenarios.
\label{fig:global}}
\end{figure*}

Electrochemical markers in the dynamics of impedances represent sharp changes of trend between $-\Delta Im$ and $+\Delta Im$ factors as discussed in Sec. \ref{sec:secFactors}: the marker A -- inside the excitation phase, the marker B -- immediately after excitation. Tests have been performed with six 30 min excitations iterated after 30 min pause. In each iteration, the excitation was increased by 5mA RMS current (6 steps from 5mA to 30mA) with two configurations of coils: 20mm length -- (78.54, 157.08, 235.6, 314.16, 392.7, 471.2) $\mu T$ for \ce{H_2O_2} setup and 10 mm length -- (157.08, 314.16, 471.2, 628.3, 785.4, 942.5) $\mu T$ for \ce{CO_2} setup to test effects of uniform and non-uniform magnetic fields. Without excitations, e.g. in the control channel 2, the impedance decreases with increasing temperature in all scenarios with/without \ce{CO_2} input, i.e. only the factor ($-\Delta Im: + \Delta t$) plays the main role, see examples in Fig. \ref{fig:global}. Markers A and B are collected in Table \ref{tab:markers} for different scenarios, expressed as timing of corresponding changes of trends, see explanations in Figs. \ref{fig:global3}, \ref{fig:global4} for \ce{^+_-A},\ce{^+_-B} notations.

\begin{table}[htp]
\centering
\caption{\small Table of electrochemical markers A and B (expressed in min after begin/end of excitation, accuracy of estimation $\pm 1$ min), notations: 'OS,7/12' -- open surface scenario with 7/12 ml fluids; '\ce{H_2O_2},0.06,air' -- \ce{H_2O_2} scenario with 0.06ml 3\% \ce{H_2O_2} in 12ml pure \ce{H_2O} in open air conditions; 'no CO2,15' -- closed containers with 15 ml \ce{H_2O}.\label{tab:markers}}
\fontsize {8} {9} \selectfont
\begin{tabular}{
p{1.5cm}@{\extracolsep{3mm}}
p{0.25cm}@{\extracolsep{3mm}}
p{0.25cm}@{\extracolsep{3mm}}
p{0.25cm}@{\extracolsep{3mm}}
p{0.25cm}@{\extracolsep{3mm}}
p{0.25cm}@{\extracolsep{3mm}}
p{0.25cm}@{\extracolsep{3mm}}
p{0.25cm}@{\extracolsep{3mm}}
p{0.25cm}@{\extracolsep{3mm}}
p{0.25cm}@{\extracolsep{3mm}}
p{0.25cm}@{\extracolsep{3mm}}
}\hline
              & \multicolumn{2}{c}{10mA} & \multicolumn{2}{c}{15mA} & \multicolumn{2}{c}{20mA} & \multicolumn{2}{c}{25mA} & \multicolumn{2}{c}{30mA} \\\hline
scenario           & A     & B         &   A   &  B  &   A & B &    A & B &   A & B \\\hline \hline
\multicolumn{11}{c}{no \ce{CO_2} input during 12 hours}\\ \hline
\ce{H_2O_2},0.03     &       & \ce{^-_+ 4}& \ce{^+_- 4}  & \ce{^-_+ 4}  &   \ce{^+_- 4} & \ce{^-_+ 4} & \ce{^+_- 4} & \ce{^-_+ 4} & \ce{^+_- 4} & \ce{^-_+ 4} \\
no CO2,15               &       &           &  &   &  &  &  &  &  & \\\hline
\multicolumn{11}{c}{no \ce{CO_2} input during 48 hours}\\ \hline
\ce{H_2O_2},0.03     &       & \ce{^-_+ 4}& \ce{^+_- 4}  & \ce{^-_+ 4}  &   \ce{^+_- 4} & \ce{^-_+ 4} & \ce{^+_- 4} & \ce{^-_+ 4} & \ce{^+_- 4} & \ce{^-_+ 4} \\
no CO2,15                &       &           & \ce{^-_+ 15} & \ce{^-_+ 15}  & \ce{^-_+ 15} & \ce{^-_+ 15} & \ce{^-_+ 15} & \ce{^-_+ 15} & \ce{^-_+ 15} & \ce{^-_+ 15} \\\hline
\multicolumn{11}{c}{open surface scenario, 1200-1500ppm \ce{CO_2}}\\ \hline
CO2 slow,12         &       &           &     & \ce{^+_- 6} & \ce{^+_- 6} & \ce{^+_- 6} & \ce{^+_- 6} & \ce{^+_- 6} & \ce{^+_- 6} & \ce{^+_- 6} \\
OS,12               &       &\ce{^+_- 10} &    & \ce{^+_- 12} &    & \ce{^+_- 15} &    & \ce{^+_- 10} &    & \ce{^+_- 10} \\
OS,12,air          &\ce{^-_- 10} & \ce{^+_- 10} & \ce{^-_- 10} & \ce{^+_- 10} & \ce{^-_- 10} & \ce{^+_- 10} & \ce{^-_- 10} & \ce{^+_- 10} & \ce{^-_- 10} & \ce{^+_- 10} \\
OS,7                &       &  &  &  &  & \ce{^+_- 15} &  & \ce{^+_- 15} &  & \ce{^+_- 15} \\\hline
\multicolumn{11}{c}{\ce{H_2O_2} scenario, 1200-1500ppm \ce{CO_2}}\\ \hline
\ce{H_2O_2},0.03    &       & \ce{^-_+ 4}&   \ce{^+_- 4}   & \ce{^-_+ 4}  &   \ce{^+_- 4} & \ce{^-_+ 4} & \ce{^+_- 4} & \ce{^-_+ 4} & \ce{^+_- 4} & \ce{^-_+ 4} \\
\ce{H_2O_2},0.06    &       & \ce{^-_+ 5}&   \ce{^+_- 5}   & \ce{^-_+ 5}  &   \ce{^+_- 5} & \ce{^-_+ 5} & \ce{^+_- 5} & \ce{^-_+ 5} & \ce{^+_- 5} & \ce{^-_+ 5} \\
\ce{H_2O_2},0.06,air    &       & \ce{^-_+ 5}&     & \ce{^-_+ 5}  &   & \ce{^-_+ 5} &  & \ce{^-_+ 5} &  & \ce{^-_+ 5} \\
\ce{H_2O_2},0.09    &       & \ce{^-_+ 6}&   \ce{^+_- 6}   & \ce{^-_+ 6}  &   \ce{^+_- 6} & \ce{^-_+ 6} & \ce{^+_- 6} & \ce{^-_+ 6} & \ce{^+_- 6} & \ce{^-_+ 6} \\
\hline
\end{tabular}
\end{table}

To avoid influence of temperature on results, Figs. \ref{fig:global5} and \ref{fig:global6} compare exactly the same heating cases with and without \ce{CO_2} input for \ce{H_2O_2} and open surface scenarios. We observe different slopes of impedance dynamics (reactivity of ionic production) depending only on \ce{CO_2} input, where all other factors are equal. This confirms the hypothesis expressed in Sec. \ref{sec:secFactors} about different reactivity of (\ref{eq:carbonicAcid}) triggered by weak excitations.

\begin{figure}[ht]
\centering
\subfigure[\label{fig:spectralDynamics1}]{\includegraphics[width=.49\textwidth]{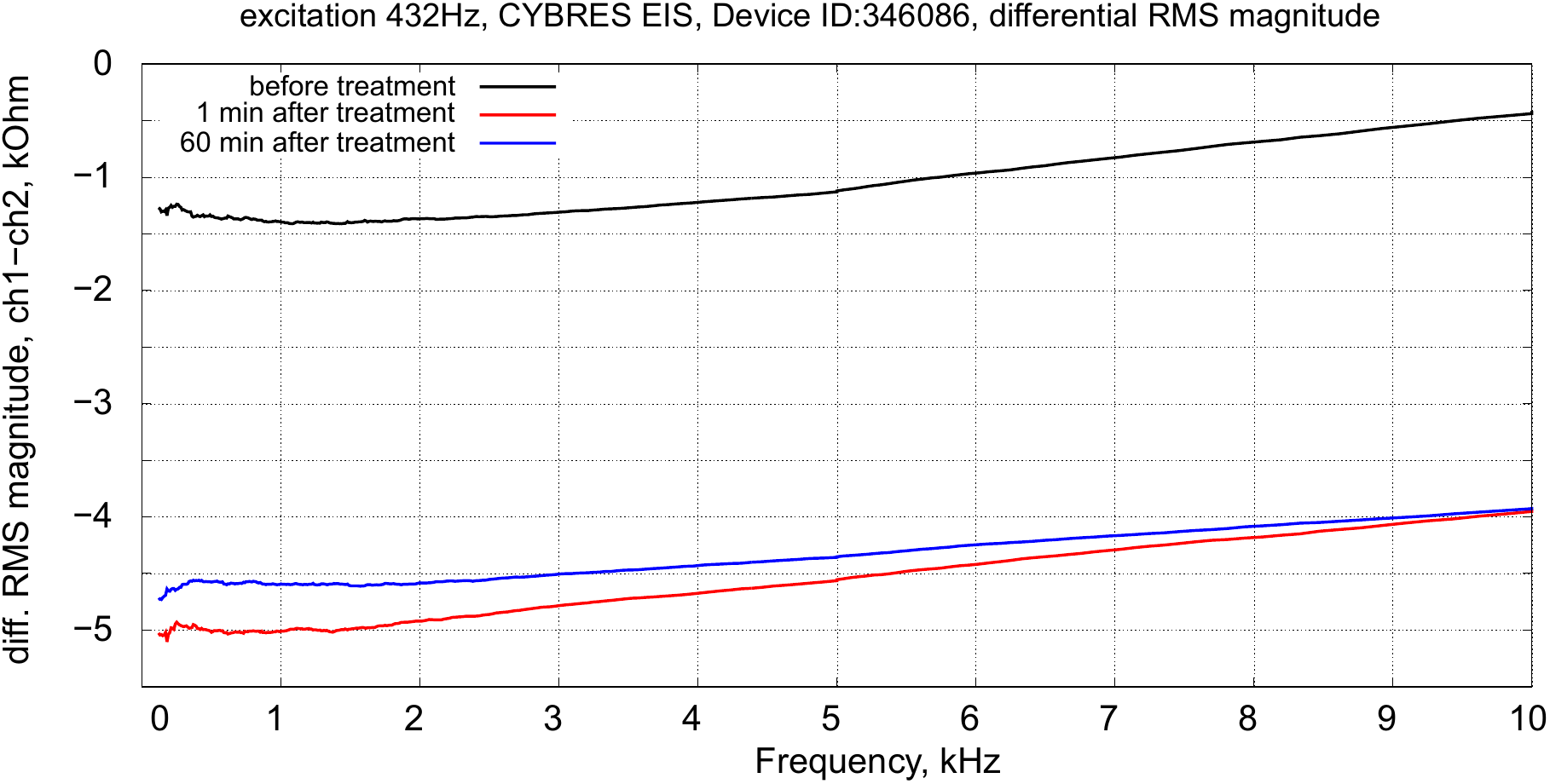}}
\subfigure[\label{fig:spectralDynamics2}]{\includegraphics[width=.49\textwidth]{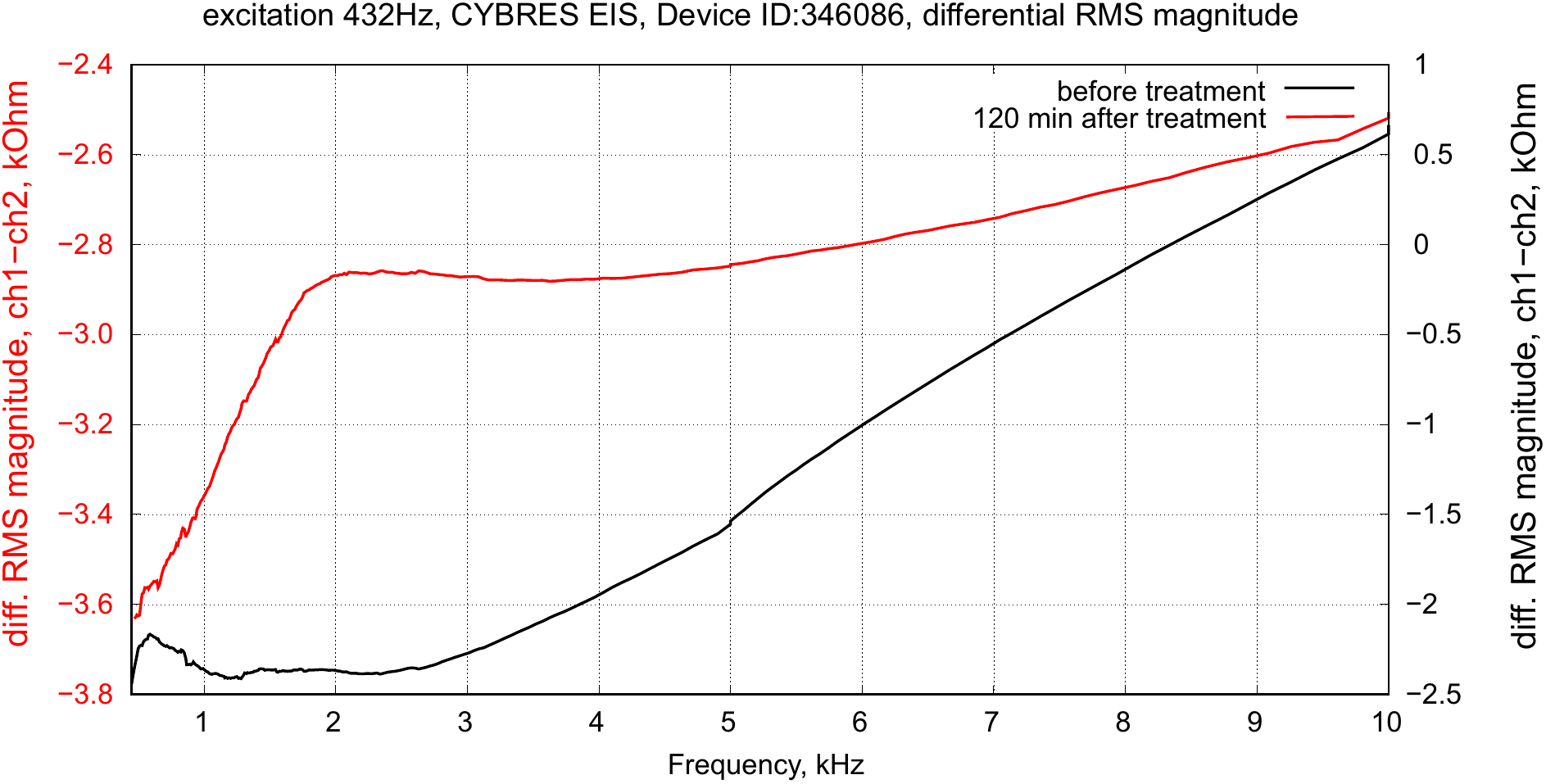}}
\subfigure[\label{fig:spectralDynamics4}]{\includegraphics[width=.49\textwidth]{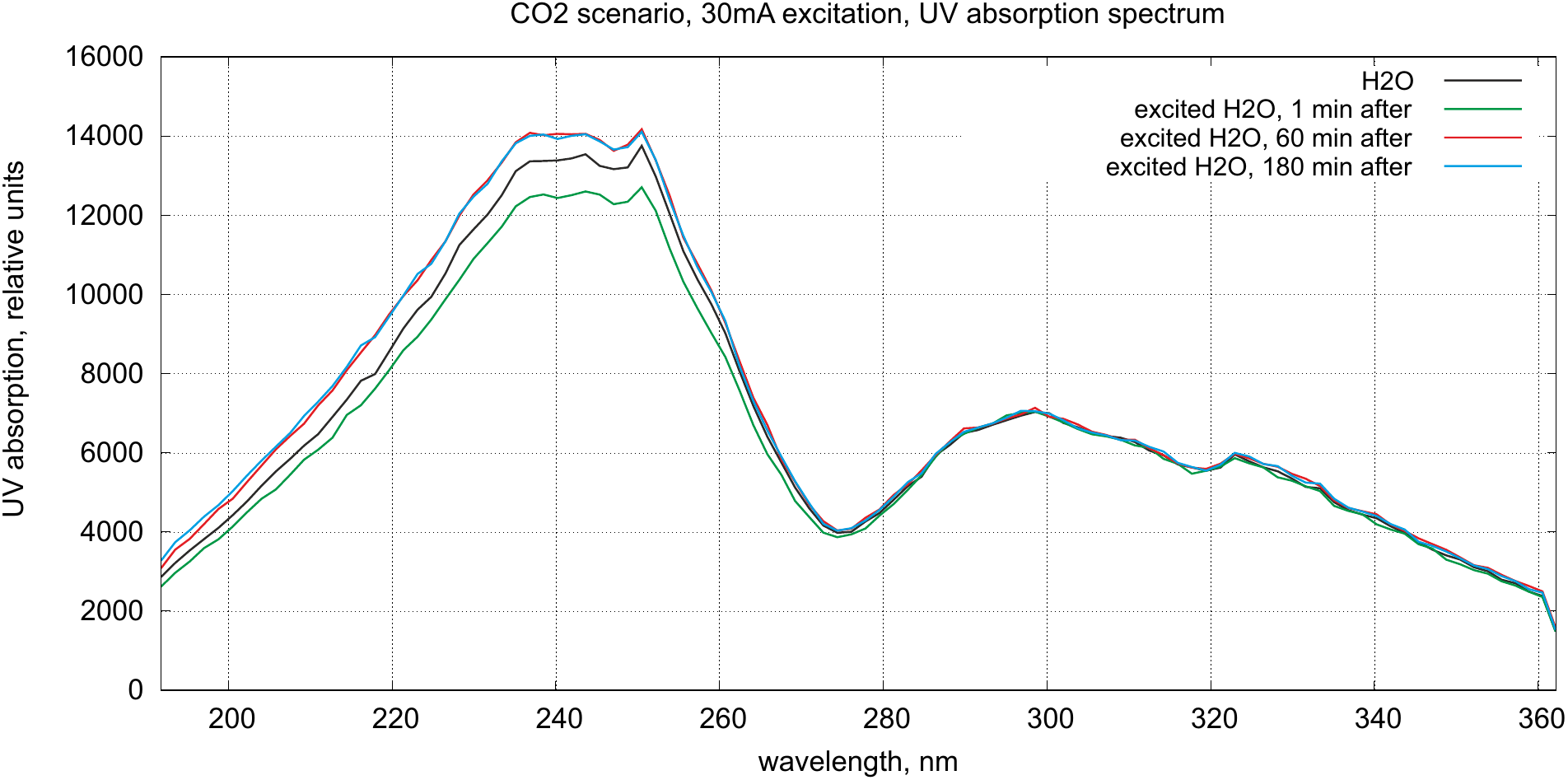}}
\caption{\small Spectra of EIS and UV absorption before and after excitation in open-\ce{CO_2} scenario: \textbf{(a,b)} differential EIS spectra; \textbf{(c)} UV absorption spectra.
\label{fig:spectralDynamics}}
\end{figure}

The temporal dynamics of impedances, shown in Figs. \ref{fig:controlImpDynamics}, \ref{fig:fastCO2}, \ref{fig:viscosity}, is also reflected in their spectral characteristics, primarily in low frequency area and UV 190-360nm range. Moreover, the fast EIS (analysis of electrochemical noise), EIS (frequency-response analysis) and UV absorption applied to the same samples before and after treatment, see Fig. \ref{fig:spectralDynamics} and Fig. \ref{fig:fastEIS} for fast EIS, demonstrate a similar character -- substantial changes immediately after excitation, and slow dynamics after this. EIS dynamics after excitation has a peak at 2kHz, see Fig. \ref{fig:spectralDynamics2}.

\section{Discussions}
\label{sec:discussion}

Based on measurements with and without access to the \ce{CO_2} atmosphere -- no significant effects without \ce{CO_2} input -- we can conclude that natural formation of ROS (natural oxygen pathway) in the used water does not play any essential role in short time interval of 180-240 minutes.

Experiments with \ce{CO_2} input have been performed in four different scenarios. Most evident results are obtained in the 'fast-\ce{CO_2}' and 'open-surface electrodes' scenarios. In the fist case, the reactions of \ce{CO_2} dissolving (\ref{eq:carbonicAcid})-(\ref{eq:carbonicAcidIonsFurther}) have a dominant character and excitation of experimental samples by magnetic field leads to a faster degradation of impedance (higher ionic reactivity), comparing to control samples without excitation. In experiments with surface tension, additional influencing factor leads to better differentiation of experimental  channel than in other experiments. Decreasing \ce{CO_2} concentration and intensity of magnetic field causes less expressed results.

Excitation by magnetic field, both in single and double (Helmholtz) solenoid configurations, introduces thermal disturbances $\Delta t$ 0.1-0.15C during 30 min at maximal excitation by 25-30mA RMS current (392.7--471.2$\mu T$ and 785.4--942.5$\mu T$ magnetic fields depending of parameters of solenoid). In fact, any energetic excitation has a thermal factor. However, the small thermal fluctuations of impedance predicted by (\ref{eq:t}) can explain neither the observed dynamics nor their inverse direction. Temperature variations with and without excitation cause different electrochemical behaviour due to ($-\Delta Im: + \Delta t$) and ($+\Delta Im: Rev$) factors and appearance of markers A and B. In experiments with equal heating dynamics, the intensity of excitation affects the ionic reactivity (slope of impedance dynamics) in regard to \ce{CO_2} ionic production as shown in Figs. \ref{fig:global5} and \ref{fig:global6}.

Due to simplicity of \ce{CO_2} dissolving on the initial stage of forming $HCO_3^-$, $H_3O^+$, $CO_3^{2-}$ ions, we argue in favour of the expressed hypothesis that the reaction (\ref{eq:carbonicAcid}) has different reactivities with/without magnetic field, which lead to forwards/reverse reactions (\ref{eq:carbonicAcid})-(\ref{eq:carbonicAcidIonsFurther}) and switching between up and down trends of impedance dynamics.

Effect of small mechanical influences as reported e.g. in \cite{ijms21218033} is also observable in our experiments. Handling of samples with comparable initial impedance after excitation but without \ce{CO_2} input provides a proportional drop of impedance, whereas the \ce{CO_2} input generates disproportionate changes after handling. To some extent, controllable mechanical stress can be utilized as a catalyst for these experiments.

These results can be considered from energy excitation levels. The oxygen pathway requires a relatively high excitation energy: Domrachev, as mentioned in \cite{Voeikov06}, used 30W/0.5kW 2-10GHz emission ($>400$J/mL) for EM treatments and generation of \ce{H_2O_2} in water. Authors in \cite{Gudkov11} excited water samples by 1264 nm laser with 1.5J (0.15J/mL), photons emit about $10^{-9}$J (effects does not appear at 0.95 eV excitation). In \cite{Shcherbakov20} authors confirm that ROS effects are not observable at mT magnetic field excitation, first measurements are done with 1T that corresponds to $10^3$ J. In this work, the intensity of magnetic field $70-700\mu T$ in performed experiments corresponds to the stored energy $10^{-7}$-$10^{-8}$J or $10^{-9}$J/mL that is considerable less than the excitation level of oxygen pathway. Pershin expressed the minimal excitation level for spin conversion at $\approx 10^{-27}$J \cite{Pershin08}, \cite{Pershin15Nano}.

To explore the induced oxygen pathway by adding a small concentration of hydrogen peroxide, we observe a completely different ionic dynamics after treatment. These reactions are briefly described e.g. in \cite{Belovolosova20} -- induced by treatment excited states of oxygen consume ions and create non-ionic \ce{H_2O_2} that first decreases the impedance. After this, \ce{H_2O_2} dissociates to \ce{HO_2^-} and \ce{H^+} that slowly increases the impedance. This two-phases-dynamics is visible in Fig. \ref{fig:h2o2}. The low energy excitation level for \ce{H_2O_2} generation is of interest. Two candidates -- the reactions (\ref{eq:domrachev}) and (\ref{eq:polack}) -- have a non-ionic form and require different instrumentation for detection (see e.g.  \cite{PMID:34855917}). Since molecular hydrogen is involved in the oxygen pathway
\begin{equation}
\label{eq:candidates}
H^\bullet + H^\bullet \rightarrow H_2, ~~~ HO^\bullet + H_2 \rightarrow H_2O + H^\bullet,
\end{equation}
it is open to what extent spin isomers of \ce{H_2}, \ce{H_2O} and excited states of oxygen influence the oxygen pathway and can explain the low-energy excitation of \ce{H_2O_2} solution.

The oscillation dynamics of different electrochemical parameters (inclusive temperature of fluids) has different reasons: reactions with \ce{HCO_3^-} ions \cite{Voeikov10}, exothermic/endothermic reactions in \ce{CO_2} and \ce{O_2} pathways, or changing of energetic levels in spin conversion mechanisms. Several theoretical and experimental works indicate a possibility of spin conversion based on reorientation of spins without energetic excitation, such as proton tunnelling that translates own spin states to neighbour molecules \cite{Konyukhov11} or micro- and nano-particles of different oxides in a normal atmosphere that changes the spin conversion period \cite{Konuchov95}. Such spin-reorientations should have energetic consequences, e.g. in the form of temperature fluctuations.

\section{Conclusion}
\label{sec:conclusion}

Concluding this work, we observed evident and characteristic changes of electrochemical dynamics in \ce{CO_2} and \ce{O_2} pathways that correlate with their chemical descriptions in assumption of different reactivity triggered by weak magnetic field. Electrochemical markers are specific for each setup, the slope of impedance degradation (reactivity of ionic production) corresponds to intensities of magnetic field. The \ce{CO_2} input represents a key factor that enables changes of ionic reactivity.

Experiments demonstrated the same phenomenology as already reported in other publications, e.g. \cite{Konuchov95}, \cite{ijms21218033} -- exposure to a normal atmosphere and slight mechanical influences (e.g. handling of samples for spectroscopy) increase the level of results -- differences between control and experimental samples in the same conditions (beside magnetic field). Since the \ce{CO_2} is more simple than the  \ce{O_2} pathway, we do not see arguments to rejects the spin conversion hypothesis from \cite{Pershin15Biophysics}, \cite{Pershin12Phytosyntesis} applied to the process (\ref{eq:carbonicAcid}). Moreover, the expressed assumption in these publications about different roles of reactivity (\ref{eq:carbonicAcid}) and kinematic viscosity (surface tension) seems to be valid -- a higher level of results is observed in the setup that involves surface tension as independent factor.

In regard of \ce{O_2} pathway, we cannot make any conclusions because all molecules having spin isomers and different excited spin configurations -- hydrogen, oxygen and water -- are involved in reactions. The level of excitation is much lower than typically used for generation of singlet oxygen, this opens questions about energy conversion mechanisms of this pathway, as indicated by (\ref{eq:domrachev}) or (\ref{eq:polack}). From practical perspective, two scenarios -- the 'open-surface electrodes' and adding \ce{H_2O_2} solution -- demonstrated well reproducible results and can be used in further experiments.

The role of signal waveforms and frequencies used for generation of magnetic field as well as other excitation approachers are not explored in this work. Due to many possibilities and a need of statistic evaluation, these topics require further attention in a separate work.

Experiments also indicated dependencies to yet unknown factors such as internal spin-reorientation mechanisms or external micro- and nano-particles in air and water that influence the results. It is evident that excitation by weak magnetic field triggers different ionic reactivity, but exact dynamics depends on multiple factors. Electrochemical impedance spectroscopy is sensitive enough for characterization and monitoring of spin conversion mechanisms and is affordable for developing different ionic-based sensing technologies.

\section{Acknowledgement}

This work was performed within the project WATCHPLANT (project grant 101017899) funded by European Commission, C.3 -- Future and Emerging Technologies. Author thanks M.Trukhanova, V.Zhigalov, V.Pacheluga and V.Voeikov for fruitful discussions about spin conversion mechanisms and energy transformation pathways in liquid water.

%\bibliographystyle{unsrt}
%\bibliography{../bib-para,../own_bibl_sk,../bibl_sk}

\section{Supplementary materials}

\begin{figure}[ht]
\centering
\subfigure[\label{fig:openOpen7}]{\includegraphics[width=.49\textwidth]{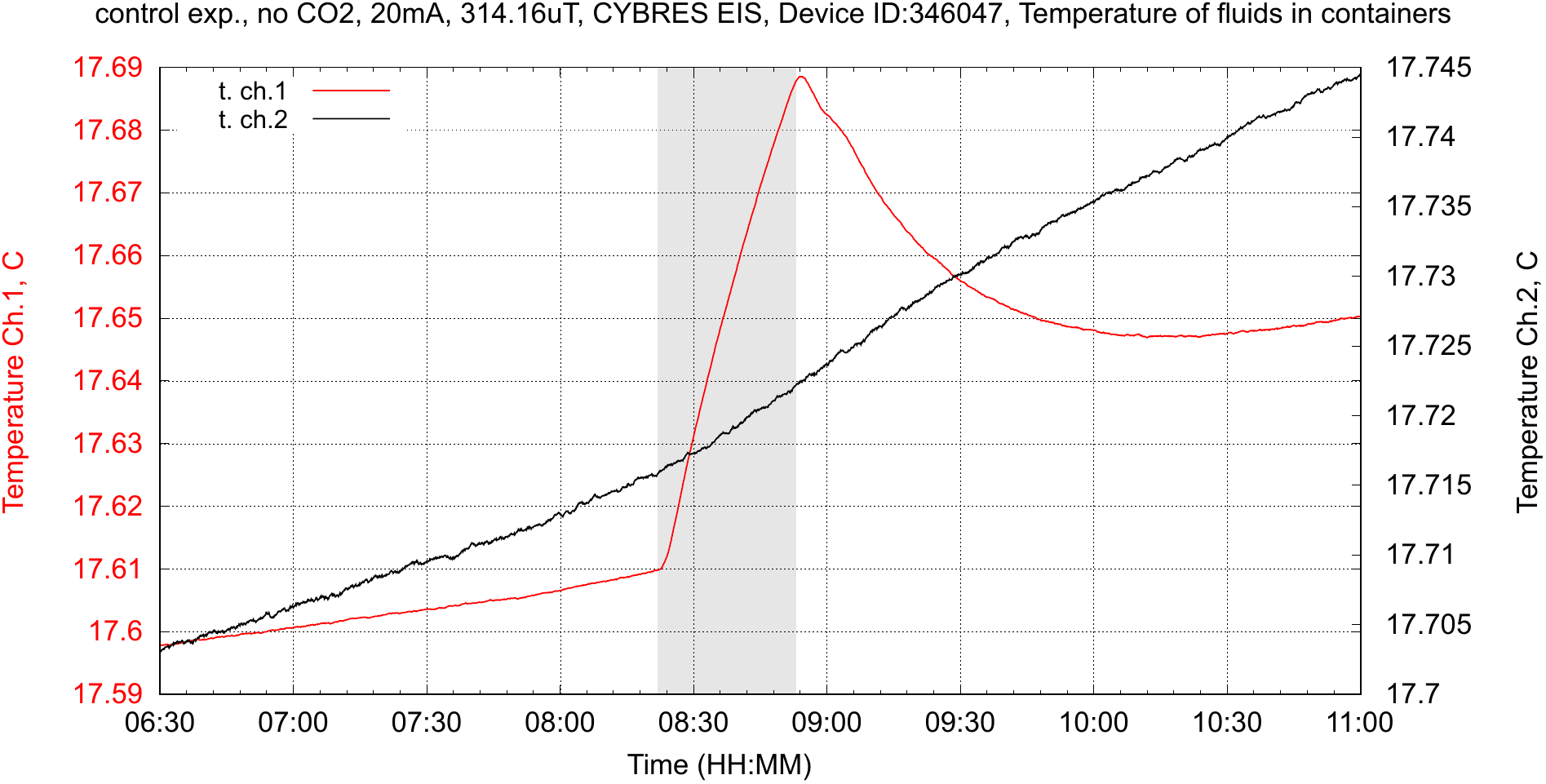}}
\subfigure[\label{fig:openOpen4}]{\includegraphics[width=.49\textwidth]{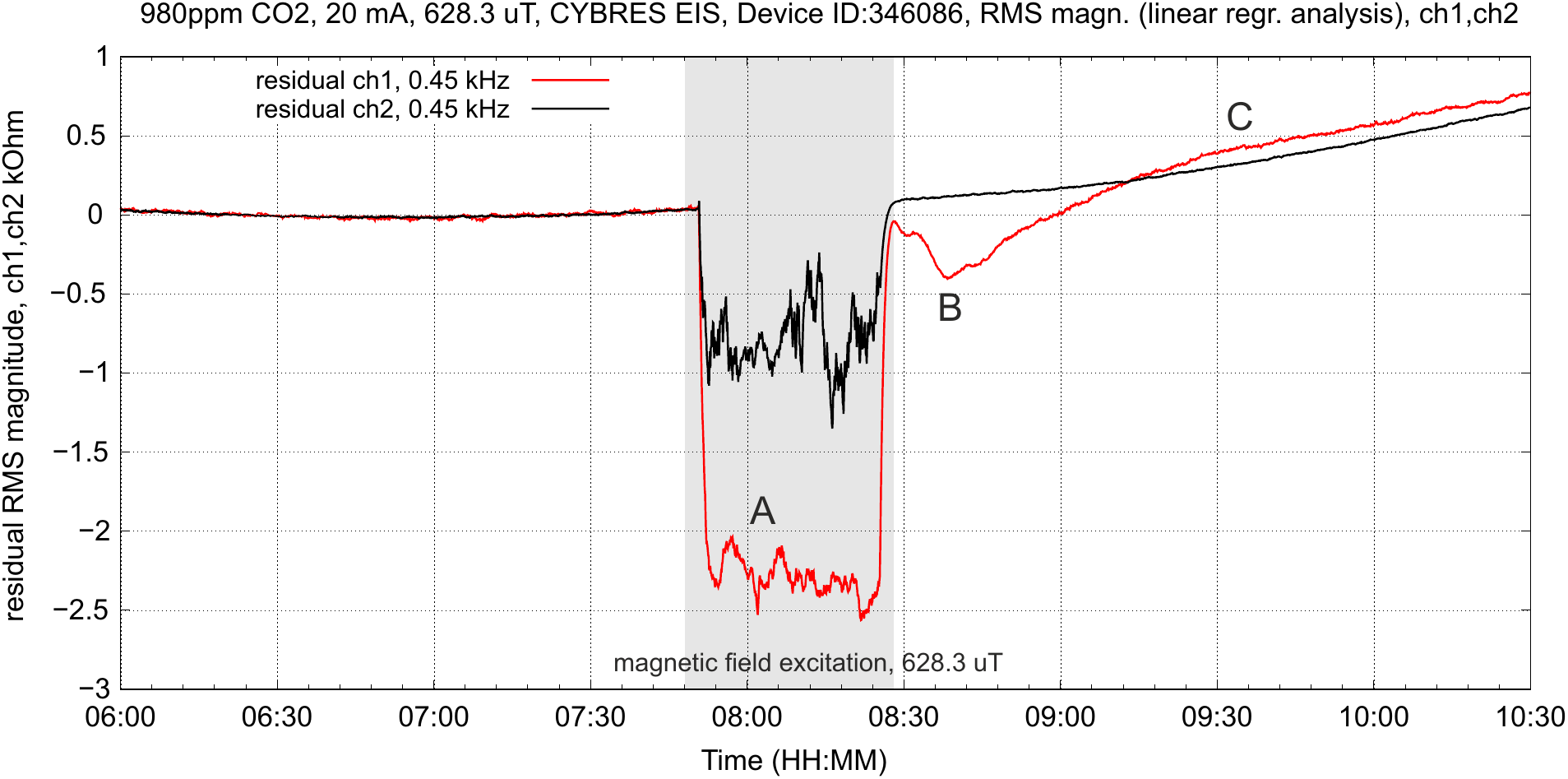}}
\subfigure[\label{fig:openOpen5}]{\includegraphics[width=.49\textwidth]{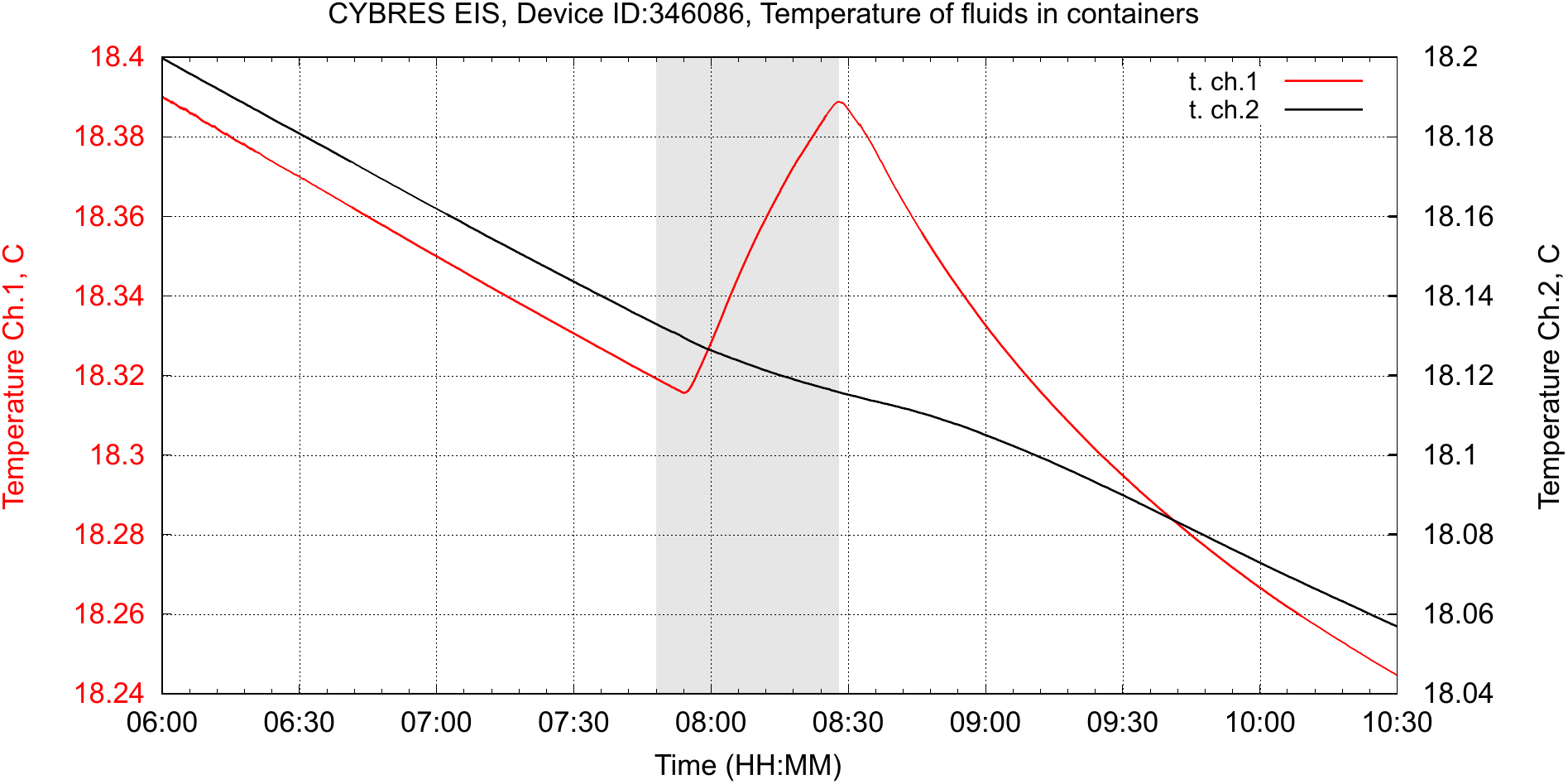}}
\subfigure[\label{fig:openOpen3}]{\includegraphics[width=.5\textwidth]{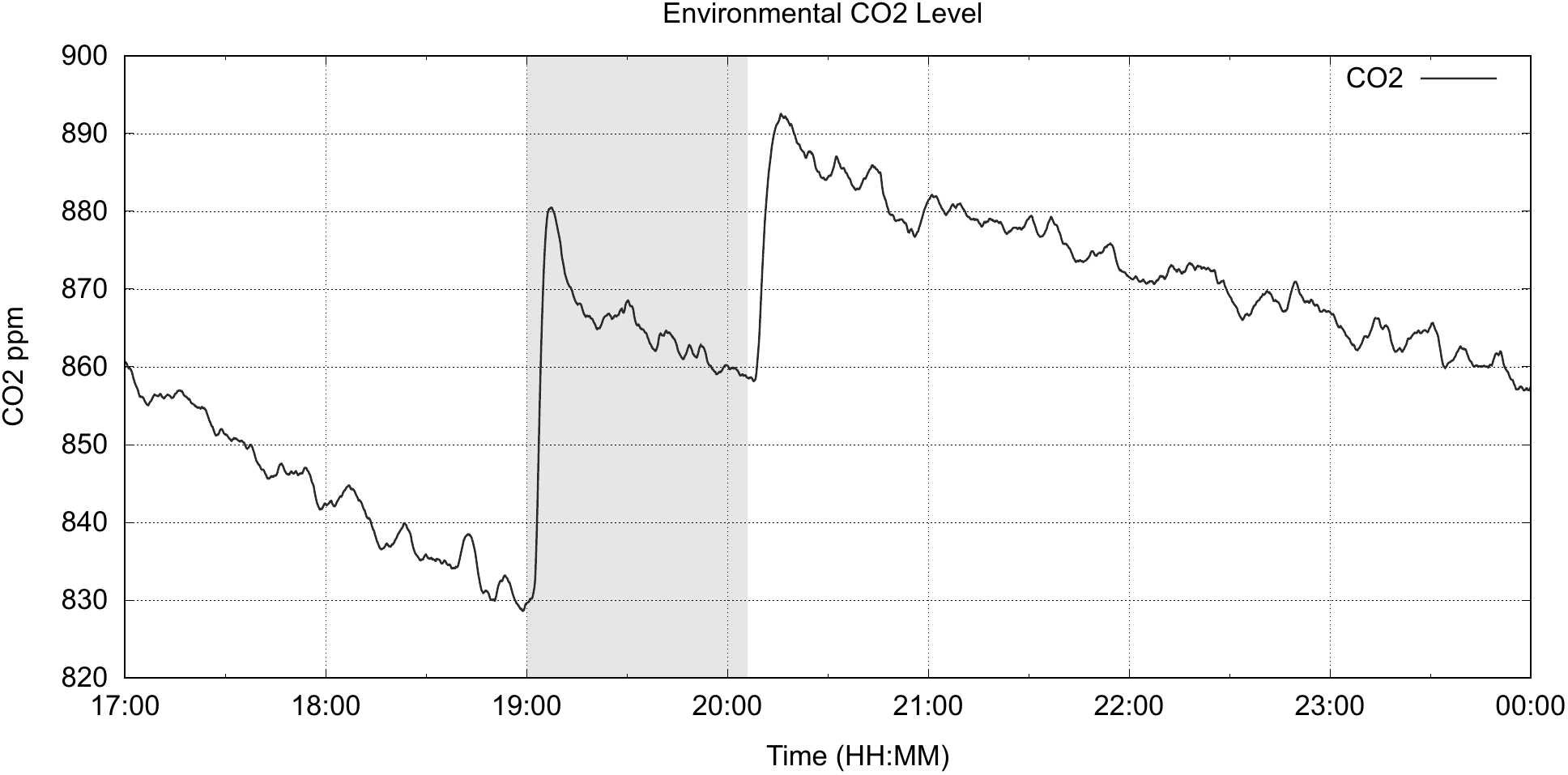}}
\caption{\small Supplementary images to Fig. \ref{fig:controlImpDynamics}: the 'low-\ce{CO_2} scenario', measurements of impedances (linear regression), temperature and \ce{CO2}: \textbf{(a)} dynamics of temperature in the control measurements without access to \ce{CO_2} at 314.16$\mu T$ with 20mA RMS current; {\textbf{(b)}} Experiment with access to \ce{CO_2} at 980ppm, 628.3$\mu T$ with 20 mA RMS current, no handling of samples and \textbf{(c)} dynamics of temperature in this measurement; {\textbf{(d)}} Environmental \ce{CO_2} close to water samples during this experiment with access to \ce{CO_2} in open-air conditions, about 500$\mu T$ with handling of samples.
\label{fig:controlImpDynamicsSup}}
\end{figure}

\end{document}